\documentclass[pre,aps,twocolumn,showpacs,showkeys]{revtex4}
\usepackage{ifpdf}
\pdfoutput=1
\usepackage{amsfonts}
\usepackage{amsmath}
\usepackage{amssymb}
\usepackage{graphicx}
\usepackage{bm}

\usepackage[english]{babel}
\usepackage{amsmath}
\usepackage[T1]{fontenc}

\begin{document}

\title{Solvent mediated forces in critical fluids}

\author{Pietro Anzini}
\email{panzini@uninsubria.it}
\affiliation{Dipartimento di Scienza e Alta Tecnologia, Universit\`a dell'Insubria, Via Valleggio 11, 22100 Como, Italy}

\author{Alberto Parola}
\email{alberto.parola@uninsubria.it}
\affiliation{Dipartimento di Scienza e Alta Tecnologia, Universit\`a dell'Insubria, Via Valleggio 11, 22100 Como, Italy}

\begin{abstract}
The effective interaction between two planar walls immersed in a fluid is investigated by use of Density Functional Theory 
in the \hbox{super-critical} region of the phase diagram. A hard core Yukawa model of fluid is studied with special attention
to the critical region. 
To achieve this goal a new formulation of the Weighted Density Approximation coupled with the Hierarchical Reference Theory, 
able to deal with critical long wavelength fluctuations, is put forward and compared with other approaches. 
The effective interaction between the walls is seen to 
change character on lowering the temperature: The strong oscillations
induced by layering of the molecules, typical of the depletion mechanism in hard core systems, 
are gradually smoothed and, close to the critical point, a long range attractive tail emerges
leading to a scaling form which agrees with the expectations based on the critical Casimir effect. 
Strong corrections to scaling are seen to affect the results up to very small reduced temperatures.
By use of Derjaguin approximation, this investigation has natural implications for the 
aggregation of colloidal particles in critical solvents. 
\end{abstract}

\pacs{05.70.Jk, 61.20.Gy, 64.60.F-}
\keywords{Fluctuation-induced interaction, Critical Casimir effect, Density Functional Theory (DFT), Inhomogeneous classical fluid}

\maketitle

\section{Introduction}
The concept of solvent mediated interaction among colloidal particles provides an invaluable 
tool for the physical understanding of the behavior of complex fluids. 
When two hard \hbox{nano-particles} are immersed 
in a molecular solvent, the simple excluded volume constraint is enough to cause
an effective attraction between them driven by purely entropic effects, 
as shown in the seminal work by Asakura and Oosawa (AO) \cite{asa_osava_54,asa_osava_58}.
The depletion mechanism has been 
extensively studied, mostly in hard sphere fluids \cite{roth_depletion_2000,ashton_wilding_roth_evans_2011}, 
and represents a very successful paradigm for the 
interpretation of colloidal aggregation. According to the AO prediction, which neglects correlations,
the strength of attraction
is proportional to the fluid density and its range coincides with the diameter of the solvent molecules. 
However, both numerical simulations and liquid state theories 
proved that the effective interaction acquires a remarkable structure when the density of 
solvent molecules is increased, due to the presence of 
coordination shells in liquids. 
Attractive interactions between solvent molecules reduce particle accumulation near the walls
giving rise to smoother solvent mediated forces \cite{louis_effective_2002}.
In the neighborhoods of the critical point of the fluid the critical Casimir effect \cite{fisher_wall_1978}
is expected to give rise to long range tails in the solvent mediated attraction between colloidal particles:
The effective force between two bodies 
acquires the range of the correlation length of the underlying critical
solvent and is described by a universal scaling law which depends on 
a very limited number of features (e.g. the geometry
of the bodies, the boundary conditions at the colloidal surfaces, the system dimensionality). 
This remarkable phenomenon
has been indeed experimentally observed by use of advanced optical
techniques in a lutidine solvent close to its critical point \cite{hertlein_direct_2008}.
Numerical simulations have been performed for two representative models of solvent: 
a \hbox{square-well} fluid and a fluid of hard spheres with three attractive patches. In both cases 
the emergence of a long range attractive tail in the effective interaction between two large 
spheres immersed in the solvent is clearly 
visible \cite{gnan_zacca_scio_symmboundaries,gnan_zacca_scio_allboundaries}.
An enhanced rate of aggregation among colloidal particles is expected to occur
as a result of critical Casimir effect, and indeed also this consequence has been directly 
observed experimentally \cite{buzzaccaro_critical_2010,buzzaccaro_enhancement_2010}.  
These investigations suggest a novel 
way to control colloidal aggregation by tuning the solvent temperature.

Critical Casimir effect is due to the confinement of order parameter fluctuations induced 
by the presence of external boundaries. As such, it governs the long range properties
of the effective interaction, while at short range excluded volume effects are still present, 
providing the essential ingredient for the occurrence of the depletion phenomenon. The interplay 
between these two quite different physical mechanisms which act at different \hbox{length-scales} has not 
been deeply investigated. Numerical simulations showed that, on approaching the critical point, 
the effective force indeed loses the short range oscillations due to the liquid layering near the 
surface of the bodies and acquires a smooth attractive 
form \cite{gnan_zacca_scio_symmboundaries,gnan_zacca_scio_allboundaries}
but no theoretical study of this effect has been attempted yet. More generally, a thorough 
investigation of the form of solvent mediated interactions in the whole phase diagram of 
a correlated fluid has not been performed. Such a study would clarify the emergence of the 
long range tails in the effective interactions on approaching the critical region, 
as well as the coexistence of the depletion mechanism, acting at short distances, and the 
critical Casimir effect, governing the physics at large distances. 

In this paper we tackle this problem by examining the solvent mediated forces between two hard walls
immersed in a hard core Yukawa fluid. 
The theoretical investigation is performed in the \hbox{super-critical} region 
in order to exclude wetting phenomena, which deserve a separate analysis. 
Preliminarily, we must choose the most appropriate theoretical tool for such a study:
A natural choice is Density Functional Theory (DFT), which proved accurate 
in the description of confined fluids. However, no specific implementation of DFT has been 
investigated in the critical region and most of the liquid state theories, 
which provide the theoretical basis for the practical formulation of DFT, give a poor 
description of the critical regime. In Section II we propose a Weighted Density Approximation (WDA) 
especially designed for being accurate also near the critical point. This theory is validated 
against other DFT prescriptions as well as available numerical simulations in several \hbox{non-critical}
states. In Section III this approach is applied to the critical region of a Yukawa fluid: 
The density profiles and the solvent mediated force between two walls are determined 
by numerical minimization of the Density Functional. The development of long range tails 
in both the density profile and the effective force is quantitatively investigated, confirming the trends 
already shown in simulations. In Section IV the universal properties of the effective 
interaction in the critical region are discussed. The emergence of scaling laws is 
investigated in the critical and \hbox{pre-critical} regime. We also show that, within our approximate
DFT formulation, the critical Casimir scaling function is deeply related to the universal 
bulk properties of the critical fluid. A simple prescription for the theoretical evaluation of 
the critical and \hbox{off-critical} Casimir scaling function is compared to the numerical results and
to available simulations in Ising systems.
Section V offers some final comments ad perspectives.

\section{Density Functional Theory}
The purpose of this work is the investigation of the solvent mediated interaction between 
two planar parallel walls immersed in a classical fluid in a wide range of temperatures and 
densities, including the critical region. 
In order to achieve this goal we need an accurate description of the
properties of a confined fluid. The most successful theoretical approach to study inhomogeneous systems
is Density Functional Theory \cite{lowen_issue_DFT}. 
Although alternative techniques, 
like integral equations or scaled particle theory, 
have been proposed \cite{henderson1992fundamental_book}, DFT is generally considered 
to be the most accurate and versatile tool for dealing with inhomogeneous systems and has been applied
in several frameworks: from the study of fluids in nanopores, to the structure of the liquid-vapor 
interface, to the theory of freezing (see e.g. \cite{simpleliquids_fourth}).

\subsection{The Weighted Density Approximation}
According to Mermin's extension \cite{mermin_65} of the Hohenberg and Kohn density functional 
theorem \cite{hohenberk_kohn}, 
the equilibrium density profile $\rho(\bm{r})$ of a confined fluid can be found by 
minimizing a suitably defined density functional $\Omega[n(\bm{r})]$ 
at fixed chemical potential $\mu$ and temperature $T$. 
$\Omega[n(\bm{r})]$ can be conveniently expressed in terms of the 
external potential $\phi(\bm{r})$ and the 
intrinsic free energy functional $\mathcal{F}[n(\bm{r})]$ as: 
\begin{equation}
\Omega[n(\bm{r})]=\mathcal{F}[n(\bm{r})]-\int \mathrm{d}\bm{r}\, n(\bm{r})
\bigl( \mu -\phi(\bm{r})\bigr). 
\label{eq:omega}
\end{equation}
The intrinsic free energy functional is exactly known only in the ideal gas limit 
\begin{equation}
\beta \mathcal{F}^{\mathrm{id}}[n(\bm{r})]=
\int \mathrm{d}\bm{r}\, n(\bm{r})\Bigl(\log \left({\Lambda}^3 n(\bm{r})\right)-1 \Bigr)
\end{equation}
(here $\beta=1/(\mathrm{k}_{\mathrm{B}}T)$ and $\Lambda$ is the thermal de Broglie wavelength),
while, in an interacting system, it is customary to separate this contribution, 
splitting $\mathcal{F}[n(\bm{r})]$ as the sum of 
the ideal and the excess term $\mathcal{F}^{\mathrm{ex}}[n(\bm{r})]$:
\begin{equation}
\mathcal{F}[n(\bm{r})]=\mathcal{F}^{\mathrm{id}}
[n(\bm{r})]+\mathcal{F}^{\mathrm{ex}}[n(\bm{r})].
\end{equation}
Several approximations for the excess part have been proposed over the years for dealing 
with specific problems.
Hard sphere fluids are successfully described by Rosenfeld's Fundamental Measure Theory (FMT) 
\cite{rosenfeld_FMT_89,fmt_review} also at bulk densities close to the solid 
transition. Even if the FMT is widely used and it its implementation is straightforward, 
at least for the planar geometry, Rosenfeld's approximation for the excess free energy holds 
in principle only for fluids of purely hard particles of any shape.
Attractive contributions in the interparticle potential are generally added as a mean field 
perturbation to the reference hard sphere excess free energy functional. 
However, such an approach gives only qualitative predictions, particularly in the critical region, 
because it does not take into account correlations arising from the attractive tail of the potential. 
Also the Weighted Density Approximations (WDA) by Tarazona \cite{tarazona_wda_sviluppo_85_pra} 
and Curtin and Ashcroft \cite{curtin_weighted-density-functional_1985}, whose prediction of 
the density profile of the hard sphere fluid are rather accurate, at least at moderate densities, 
includes the effects of an 
attractive tail in the potential only at the mean field level. Furthermore, approaches
based on a truncation of the gradient expansion of the excess free energy functional, such as the 
square-gradient approximation, 
can describe only the slowly varying part of the density profile but fail in dealing with the 
short range structure close to the wall. 

In order to describe the microscopic properties of the solvent mediated forces 
in the whole phase diagram of a fluid, including the liquid-vapor transition, 
we need an approximation of the excess free energy functional which, 
in the uniform limit, provides an accurate description of the homogeneous fluid both in the dense and in 
the critical regime. 
Unfortunately such an implementation of DFT has not been devised yet.  
A novel WDA-based approximation for the excess free energy functional based on the Hierachical Reference
Theory of fluids \cite{HRT_advphys_95} will be now introduced. 

According to the weighted density approximation, 
the excess free energy functional 
is expressed in terms of a weighted density $\bar{n}(\bm{r})$ as
\begin{equation}
\mathcal{F}^{\mathrm{ex}}[n(\bm{r})]=\int \mathrm{d}\bm{r}\, 
n(\bm{r}){\psi}^{\mathrm{ex}} \bigl(\bar{n}(\bm{r}) \bigr),
\label{eq:wda_excfreeenergy}
\end{equation}
where ${\psi}^{\mathrm{ex}}(\rho)$ is the excess free energy per particle of the 
homogeneous system evaluated at bulk density $\rho$ and the weighted density 
$\bar{n}(\bm{r})$ is written as a local average of the density profile, in terms of 
an isotropic weight function $w(r;\hat n)$:
\begin{equation} 
\bar{n}(\bm{r})=\int \mathrm{d}\bm{r}^\prime n(\bm{r}^\prime)
w\bigl(|\bm{r}-\bm{r}^\prime|;\hat{n}(\bm{r})\bigr).
\label{eq:weighted_density}
\end{equation} 
The weight function can be generally dependent on the local value of an auxiliary
reference density $\hat{n}(\bm{r})$ and has to satisfy the normalization requirement
\begin{equation}
\int \mathrm{d}\bm{r}^\prime \,w\bigl(|\bm{r}-\bm{r}^\prime|;\hat{n}(\bm{r})\bigr)=1 
\qquad \forall \, \bm{r},
\label{eq:normalization_w}
\end{equation}
to ensure that in the homogeneous limit the weighted density coincides with the actual density of the fluid. 

In order to get a reliable approximation for the excess free energy functional we have to 
choose properly the two key ingredients which characterize our WDA ansatz, 
namely the homogeneous free energy ${\psi}^{\mathrm{ex}}$ and the weight function 
$w\bigl(r;\hat{n}\bigr)$.
The only available microscopic liquid state theory which is able to 
account both for non critical and critical properties of a 
homogeneous fluid is the Hierarchical Reference Theory (HRT) \cite{HRT_advphys_95}, which 
will be therefore adopted in this work for the evaluation of the excess free energy $\psi^{ex}$ of 
the uniform fluid. 

Although the general formalism of HRT can be applied to fluids and mixtures 
with arbitrary pair interactions, quantitative results for specific models
require the closure of the exact HRT equations by introducing some approximation. 
A closure which proved remarkably accurate has been  
implemented in the case of a Hard Core Yukawa (HCY) fluid \cite{smoothcutoff_2008}, because 
the resulting HRT equations considerably simplify
by use of the known solution of the Ornstein-Zernike equation available for this interaction. 
The HCY potential is defined as the sum of a pure hard core term of diameter $\sigma$ and an 
attractive Yukawa tail of inverse range $\zeta$:
\begin{equation}
v_{\mathrm{Y}}({r}) =
-\epsilon\,\frac{\mathrm{e}^{-\zeta(r- \sigma)}}{r} \quad \quad r \geq \sigma, \\
\label{eq:def_Yukawa}
\end{equation}
where the parameter $\epsilon$, which defines the energy scale, is positive.
In the following we will investigate this model taking 
$\sigma$ and $\epsilon/\mathrm{k}_{\mathrm{B}}$ as the units of length and temperature respectively.

Having established the form of our excess functional in the homogeneous limit, we proceed with the explicit 
definition of $\mathcal{F}^{\mathrm{ex}}[n(\bm{r})]$ for general density profiles. 
We first focus our attention on the effects of the attractive part of the potential.  
It is well known that its main contribution to the internal energy is given by the Hartree term:
this circumstance has been extensively recognized in the previous treatments, where attractive
interactions were included just through such a contribution. 
We are therefore led to isolate this term in the excess free energy, by writing:
\begin{equation}
\mathcal{F}^{\mathrm{ex}}[n(\bm{r})]=\mathcal{F}^{\mathrm{ex}}_{\mathrm{R}}[n(\bm{r})]+
\mathcal{F}^{\mathrm{ex}}_{\mathrm{H}}[n(\bm{r})],
\label{eq:split}
\end{equation}
where the Hartree contribution is given by 
\begin{equation}
\mathcal{F}^{\mathrm{ex}}_{\mathrm{H}}[n(\bm{r})]=
\frac{1}{2}\int\mathrm{d}\bm{r}^\prime\int \mathrm{d}\bm{r}^{\prime\prime} \, n(\bm{r}^\prime) 
v_{\mathrm{Y}}(|\bm{r}^\prime-\bm{r}^{\prime\prime}|) n(\bm{r}^{\prime\prime}),
\end{equation}
whereas the reference term $\mathcal{F}^{\mathrm{ex}}_{\mathrm{R}}[n(\bm{r})]$, 
defined by Eq. (\ref{eq:split}), contains both the entropic contribution to the free energy, 
arising from hard core repulsion, and the correlations induced by the attractive interaction. 
Our choice is then to use WDA to represent only the \hbox{entropy-correlation} part of the 
intrinsic free energy functional,
retaining the exact description of the Hartree energy:
\begin{equation}
\mathcal{F}^{\mathrm{ex}}_{\mathrm{R}}[n(\bm{r})]=\int \mathrm{d}\bm{r}\, 
n(\bm{r}){\psi}^{\mathrm{ex}}_{\mathrm{R}}\bigl(\bar{n}(\bm{r}) \bigr),
\end{equation} 
where 
\begin{equation}
{\psi}^{\mathrm{ex}}_{\mathrm{R}}(\rho)={\psi}^{\mathrm{ex}}(\rho)-\frac{\rho}{2}
\int\mathrm{d}\bm{r}\,v_{\mathrm{Y}}(r).
\end{equation}
The form of the weight function can be determined following the strategy put forward by 
Tarazona \cite{tarazona_wda_sviluppo_85_pra}, by requiring that the two particle 
direct correlation function reduces, in the homogeneous limit, to the known form 
of the underlying bulk liquid state theory which, in our case, is the direct correlation function 
$c(r,\rho)$ predicted by HRT:
\begin{eqnarray}
\left.-\beta\frac{{\delta}^2 \mathcal{F}^{\mathrm{ex}}_{\mathrm{R}}[n]}{\delta n(\bm{r}) 
\delta n(\bm{r}^\prime)}\right|_{n(\bm{r})=\rho_b} 
\hskip -5mm &=& c(|\bm{r}-\bm{r}^\prime|,{\rho}_b)+
\beta \, v_{\mathrm{Y}}(|\bm{r}-\bm{r}^\prime|)
\nonumber \\
&\equiv& c_{\mathrm{R}}(|\bm{r}-\bm{r}^\prime|,{\rho}_b),
\label{eq_constraint_WDA_R}
\end{eqnarray}
where the last equality is a definition.
This constraint can be fulfilled only by a density dependent weight function. 
In a previous work Curtin and Ashcroft \cite{curtin_weighted-density-functional_1985} 
proposed that the weight function should depend on the local value of the weighted density 
itself, i.e. $\hat{n}(\bm{r})=\bar{n}(\bm{r})$. Unfortunately, 
even for a purely repulsive hard sphere system, 
this hypothesis leads to difficulties in the solution of the nonlinear differential 
equation for the weight function\footnote{It is precisely an Abel differential equation of the second kind,
which, in a range of densities, has no solutions satisfying the physical boundary conditions.} 
which must be imposed in order to implement the constraint (\ref{eq_constraint_WDA_R}). 
This problem can be overcome choosing a position independent auxiliary 
density $\hat{n}(\bm{r})$, as suggested by Leidl and Wagner \cite{wagner_HWDA}. 
In our application of DFT to the description of confined systems, 
we assume that the auxiliary density actually coincides with the homogeneous 
density at the same chemical potential (also referred to as {\it bulk density}), i.e. we set
\begin{equation}
\hat{n}(\bm{r})=\rho_b.
\end{equation}
It is straightforward to obtain from Eq. (\ref{eq_constraint_WDA_R}) an algebraic 
equation for the Fourier transform of the weight function:
\begin{eqnarray}
\left.\beta{\rho_b} \frac{{\partial}^2 {\psi}^{\mathrm{ex}}_{\mathrm{R}} }{{\partial} 
{\rho}^2}\right|_{\rho_b} && w^2 \bigl(q;{\rho}_b \bigr)+
2\beta\left.\frac{\partial {\psi}^{\mathrm{ex}}_{\mathrm{R}} }{\partial \rho}\right|_{\rho_b} 
w \bigl(q;{\rho}_b \bigr)+
\nonumber \\
&& + \,c_{\mathrm{R}}\bigl(q,{\rho}_b\bigr)=0,
\label{eq:w_rhoccostante_fourier_R}
\end{eqnarray}
which, at least in the cases examined in this work, always admits real solutions. 
The physical root can be determined enforcing the normalization condition (\ref{eq:normalization_w}), 
recalling that the compressibility sum rule, satisfied by the HRT direct correlation function, requires
\begin{equation}
c_{\mathrm{R}} (0;{\rho}) =\int \mathrm{d}\bm{r} \,c_{\mathrm{R}} \bigl(r;{\rho} \bigr)=  
-2\beta \frac{\partial {\psi}^{\mathrm{ex}}_{\mathrm{R}} }{\partial \rho}- 
\beta \rho\frac{{\partial}^2 {\psi}^{\mathrm{ex}}_{\mathrm{R}}}{{\partial} {\rho}^2}.
\end{equation}
\begin{figure}
\includegraphics[width=7cm,clip]{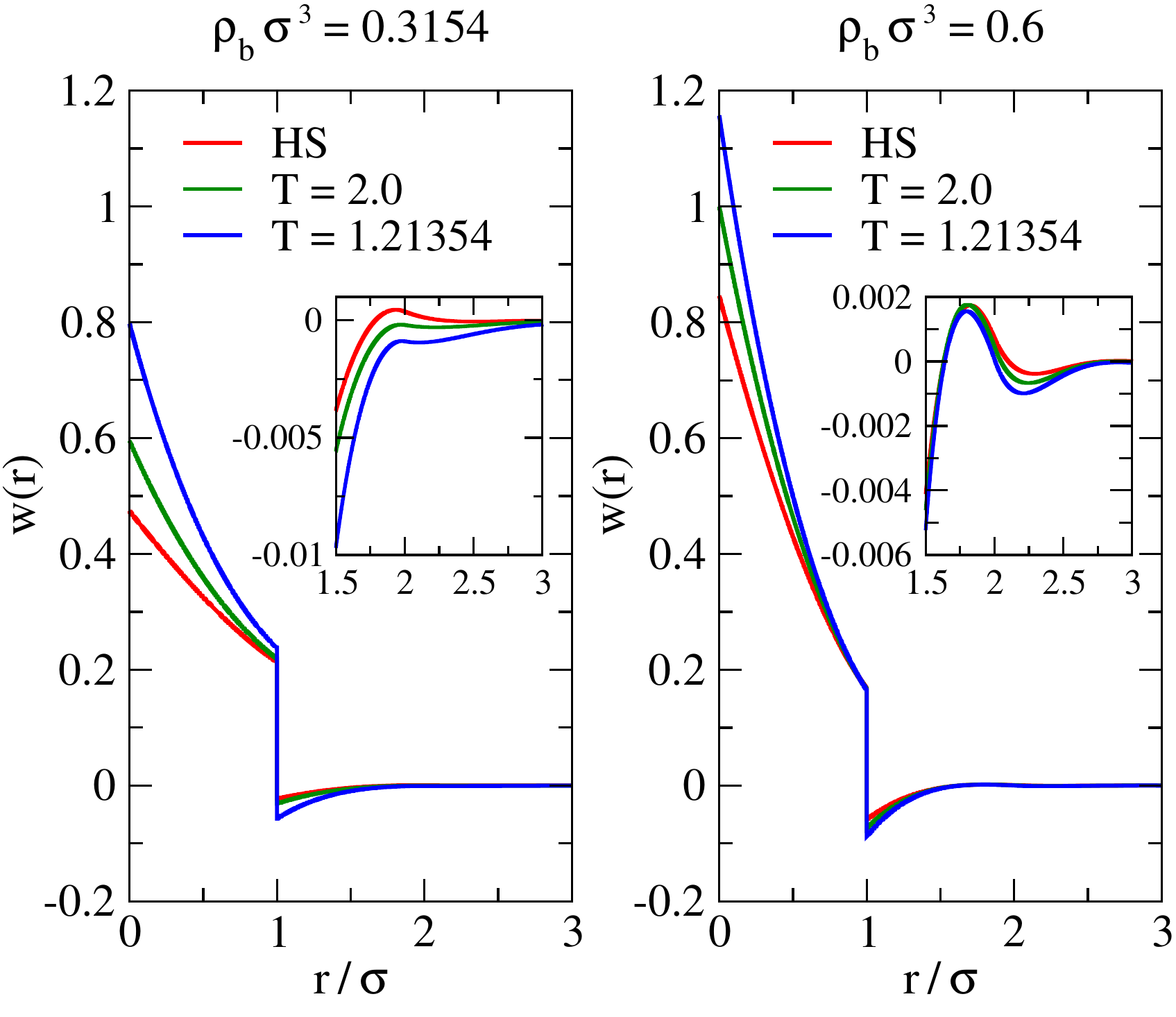}
\caption{Real space weight function at two values of bulk density for three values of the temperature.
One state shown here ($\rho_b\sigma^3 = 0.3154$, $T=1.21354$) is very close to the critical point.
}
\label{fig:weightfunction}
\end{figure}
As in other implementations of WDA, the range of the 
weight function always remains comparable to the size of the molecules.
This feature is preserved also in the critical region, as shown in Fig. \ref{fig:weightfunction}.

Once the intrinsic density functional has been determined, the approximate equilibrium properties, 
such as the density profile $\rho(\bm{r})$ and 
the grand canonical potential $\Omega[\rho(\bm{r})]$, can be obtained 
minimizing the functional (\ref{eq:omega}) at fixed temperature and chemical potential. 
In a HCY fluid, the value of the chemical potential is related to the bulk density by:
\begin{equation}
\mu =\frac{1}{\beta}\log \rho_b+{\psi}^{\mathrm{ex}}_{\mathrm{R}}(\rho_b)+
\rho_b\left.\frac{\mathrm{d}{\psi}^{\mathrm{ex}}_{\mathrm{R}}}{\mathrm{d}\rho}\right|_{\rho_b}-
4\pi\frac{\zeta+1}{{\zeta}^2}\rho_b.
\label{eq:eq_mu}
\end{equation}

A central quantity for the present investigation is 
the force acting on the two planar hard walls in the HCY fluid. 
In this geometry, symmetry requires that all the local properties may depend on the single 
coordinate $z$, orthogonal to the two plates, placed at $z=0$ and $z=L$ respectively. 
Remarkably, if the wall separation $h$ 
is greater than $\sigma$, the force per unit surface $f$ acting on the plates, 
sometimes called \emph{solvation force}, can be expressed as a pressure 
difference \cite{jr_henderson_solvation_86}:
\begin{equation}
f(L;T,\mu)=-\left.\frac{\partial}{\partial L}
\left(\frac{\Omega^{(L)}[\rho(z)]}{\Sigma} \right)\right|_{T,\mu}-p_b(T,\mu),
\label{eq:solvation_derivata}
\end{equation}
where $\Omega^{(L)}[\rho(z)]/\Sigma$ is the grand canonical potential per unit surface 
of the fluid confined in the region $[0,L]$, determined by the minimization of the 
approximated grand canonical functional at fixed $(\mu,T)$ and $p_b(\mu,T)$ is the pressure of 
the fluid at the same values of temperature $T$ and chemical potential $\mu=\mu(\rho_bT)$.
On the other hand, when $L<\sigma$ there are no particles between the walls and the attractive 
force per unit surface acting on the walls arises uniquely from the presence of the fluid in the 
regions $z<0$ and $z>L$: 
The first term in Eq. (\ref{eq:solvation_derivata}) vanishes and the force is given by
\begin{equation}
f(L;T,\mu)=-p_b(T,\mu).
\end{equation}
The solvation force $f$ defined above is a difference of pressures and goes to zero in 
the limit $L\to\infty$. 
By means of standard functional identities it is possible to express, without any further 
approximation, the derivative of the grand potential per unit surface in terms of the contact density:
\begin{equation}
\rho^{(L)}_w\equiv \lim_{\delta \to 0^+}\rho\bigl(L-\sigma/2-\delta\bigr)=
\lim_{\delta \to 0^+}\rho\bigl(\sigma/2+\delta\bigr),
\end{equation} 
and the solvation force can be finally written as \cite{evans_surfaces_theorist}:
\begin{equation}
f(L;T,\mu)=\mathrm{k}_{\mathrm{B}}T\,\rho^{(L)}_w-p_b(T,\mu).
\label{eq:solvation_rho}
\end{equation}
We also remark that this version of WDA exactly satisfies the contact value theorem \cite{henderson_swol_contact} 
\begin{equation}
\beta p_b(T,\mu)=\lim_{L\to\infty}\rho^{(L)}_w,
\end{equation}
leading to the more suggestive identity
\begin{equation}
\beta f(L;T,\mu)=\rho^{(L)}_w-\rho^{(\infty)}_w.
\end{equation}

The above results show that, in order to obtain the force acting on the walls we 
just need to perform the minimization of the functional in the region $[0,L]$ in order to evaluate the 
contact density $\rho^{(L)}_c$. The minimization has been be carried out by a simple 
iterative (Picard) method, taking advantage of the exact implicit relation for the approximate
equilibrium density profile $\rho(z)$ given by:
\begin{equation} 
\rho(z)=\exp{\left[-\beta (u(z)-\mu)\right ]}, 
\label{eq:iteration_minimization}
\end{equation}
where the potential of mean force $u(z)$ is defined as:
\begin{equation}
u(z)= \frac{\delta\mathcal{F}^{\mathrm{ex}}_{\mathrm{R}}[n]}{\delta n(\bm{r})}\Bigg|_{\rho(z)}
+\int \mathrm{d}\bm{r}^\prime \rho(z^\prime)v_{\mathrm{Y}}(|\bm{r}-\bm{r}^\prime|).
\end{equation}
A spatial \hbox{step-size} $\Delta z=1.5\times 10^{-2} \sigma$ has been generally used in the numerical minimization, 
while for bulk densities $\rho_b\sigma^3>0.7$ and close to the critical point $\Delta z$ has been reduced up to two orders of magnitude. 
Typically, up to few thousand Picard iterations were necessary to achieve a precision of one part 
in $10^7$ for the density profile and one part in $10^{11}$ in the grand potential. 

\subsection{Validation of the method}
The minimization of the previously defined grand canonical functional 
allows to evaluate the equilibrium properties of the confined fluid. 
Within this approach, the relevant quantities can be found at every temperature and bulk density of 
interest, also in the vicinity and below the critical point of the HCY fluid. 
Most of the calculations refer to a Yukawa fluid with range 
$\zeta\sigma=1.8$, where several simulation results are available.
We performed minimizations of the functional for values of the temperature $T$ 
above the critical point ($T > T_c \sim 1.21353$) and bulk reduced densities 
$\rho_b{\sigma}^3$ up to $0.85$. 

\begin{figure}
\centering
\includegraphics[width=7cm,clip]{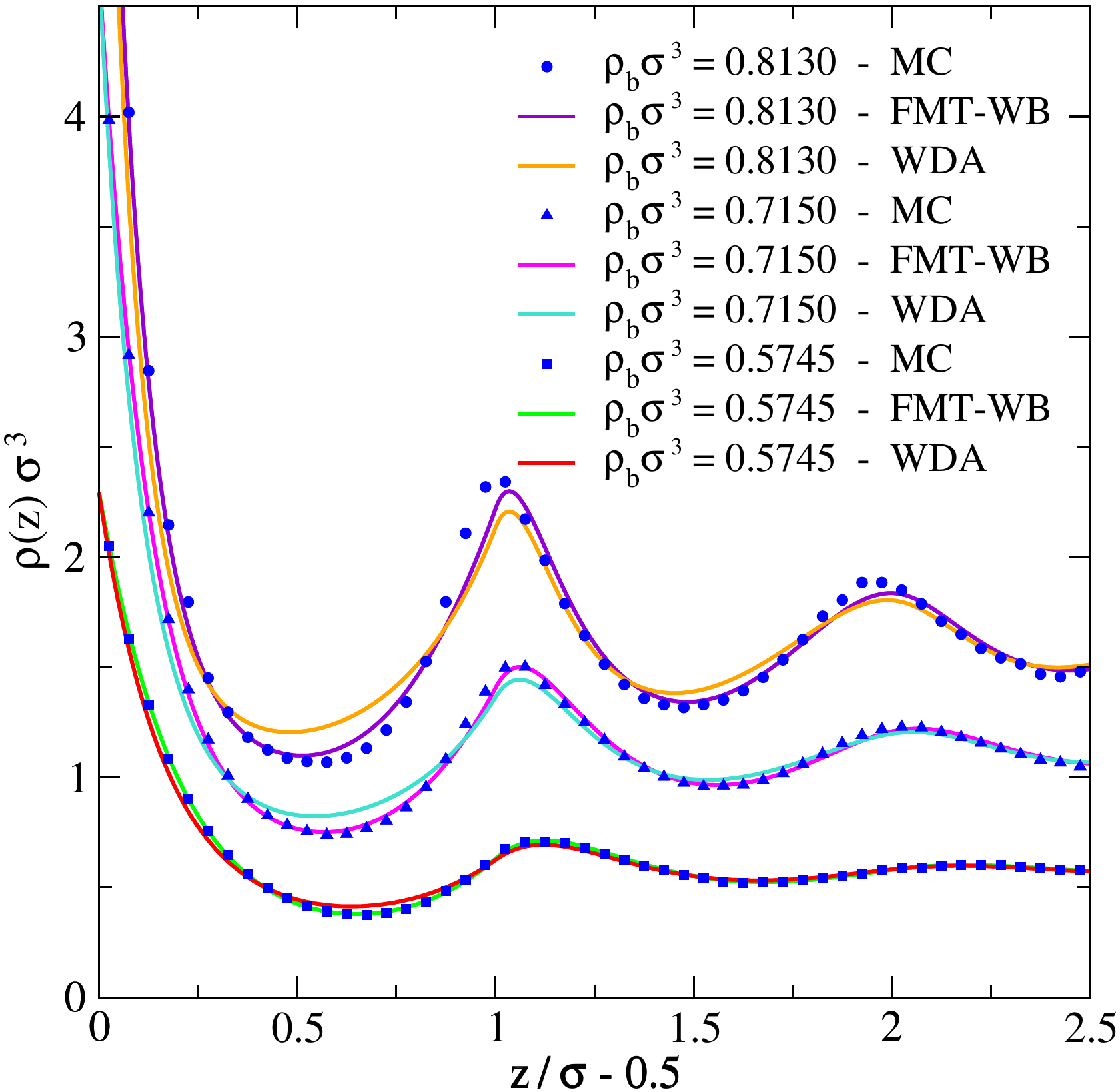}
\caption{Density profiles $\rho(z)$ for a hard sphere fluid between two hard walls 
(distance $L=16\,\sigma$) at different values of the bulk reduced density 
$\rho_b{\sigma}^3$. MC simulation data (symbols) are taken from Ref. \cite{groot_faber_vandereerden}. 
We have obtained the FMT density profiles (lines) from the minimization of the approximated 
FMT-WB functional \cite{FMT_WB}. To enhance visual clarity the density profiles at 
$\rho_b\sigma^3=0.715$ and  $\rho_b\sigma^3=0.813$ are shifted upward by $0.4$ and $0.8$ respectively.}
\label{fig:hs_simulation_fmt}
\end{figure}

\begin{figure}
\includegraphics[width=7cm,clip]{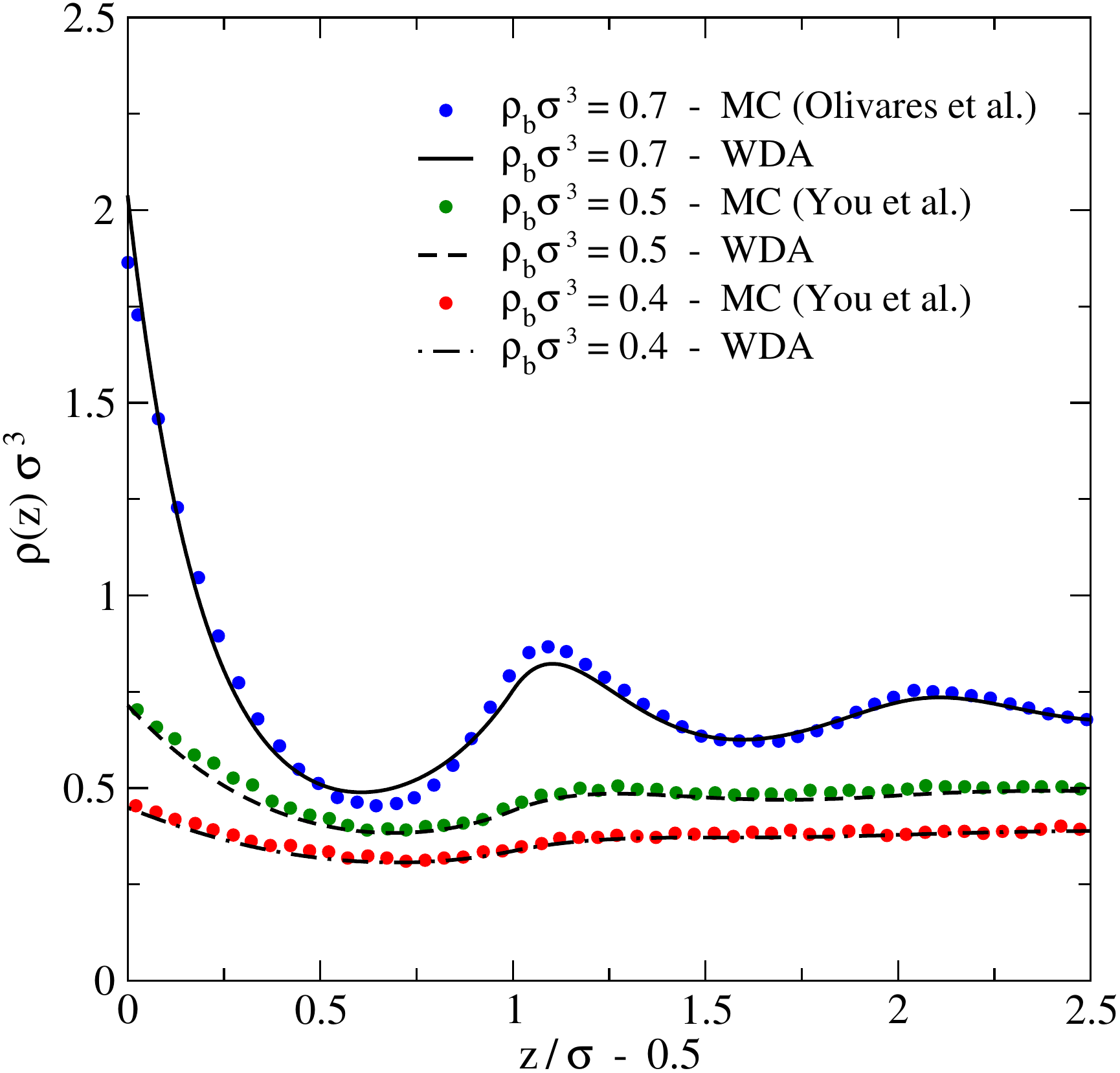}
\caption{Density profile $\rho(z)$ of a Yukawa hard sphere fluid ($\zeta \sigma=1.8$) 
at reduced temperature $T=2$ confined between two hard walls at different values of bulk 
reduced density $\rho_b{\sigma}^3$. Lines represent the predictions of the present WDA. 
Points are MC data from Ref. \cite{henderson_MCyukawa} (reduced density $0.7$) 
and Ref. \cite{You_YukawaMC} (reduced density $0.4$ and $0.5$). 
The distance between the walls is $10\,\sigma$.}
\label{fig:T2_diversirho}
\end{figure}

\begin{figure}
\includegraphics[width=7cm,clip]{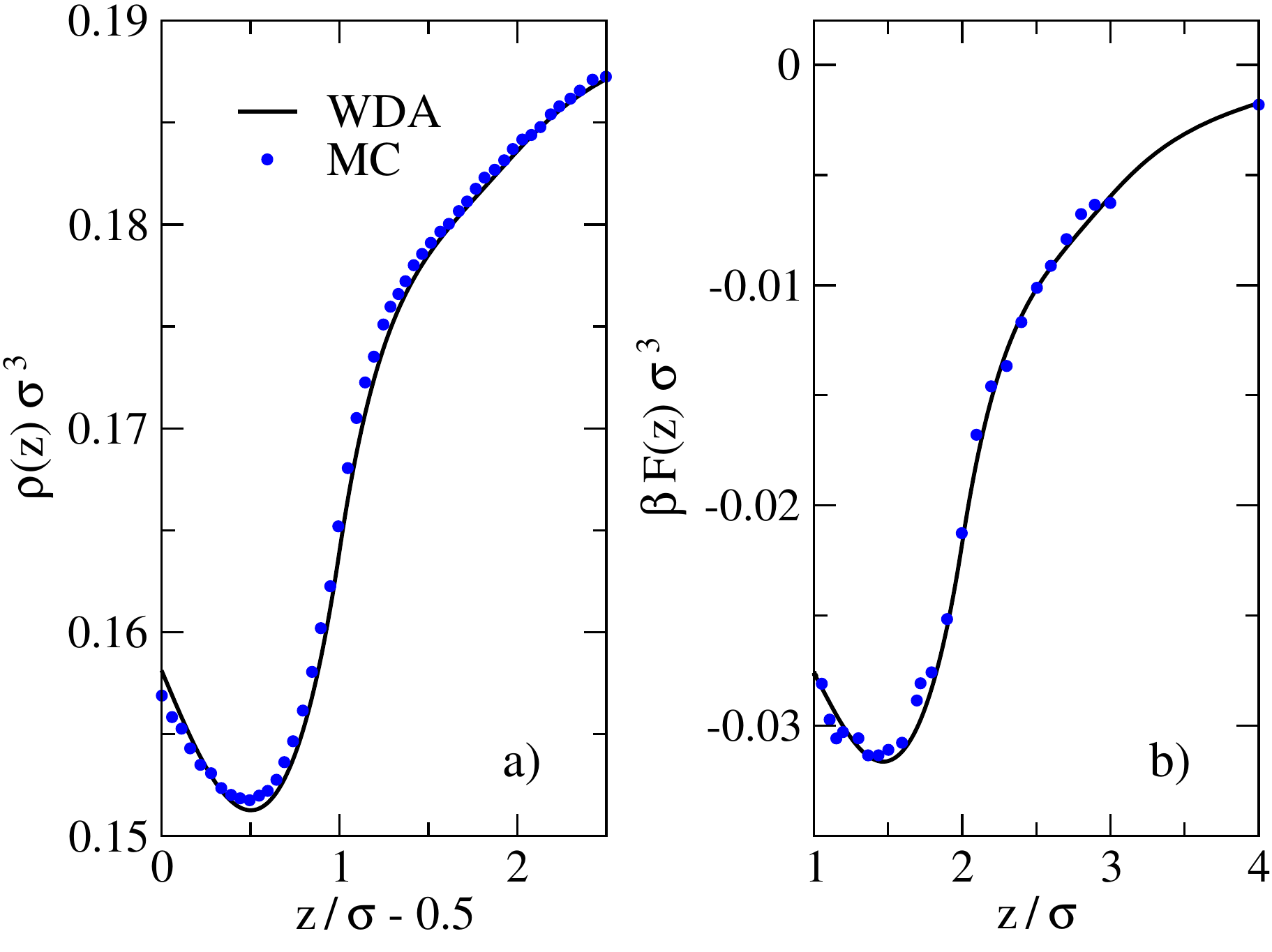}
\caption{Panel a): Density profile $\rho(z)$ of a HCY fluid at a single hard wall
($\zeta\sigma=3$) for bulk reduced density $\rho_b{\sigma}^3=0.191$
and temperature $T=1.004$. 
Lines represent the predictions of the present WDA. Points are MC data from 
Ref. \cite{louis_yukawaMC_plates}. Panel b): Force per unit surface acting between two 
infinite parallel hard walls immersed in a HCY fluid ($\zeta\sigma=3$) in the same thermodynamic 
state of panel a). Lines are the predictions of the WDA from 
Eq. (\ref{eq:solvation_rho}), whereas points represent the MD simulation 
of Ref. \cite{louis_yukawaMC_plates}.}
\label{fig:profili_forze_louis}
\end{figure}

In the high temperature limit our model reduces to a hard sphere fluid, whose properties
have been extensively investigated by numerical simulations.  
Figure \ref{fig:hs_simulation_fmt} shows the density profiles $\rho(z)$ 
of a hard sphere fluid near a hard wall at two different values of $\rho_b$.
The agreement of the WDA prediction
with the Monte Carlo (MC)  data of Ref. \cite{groot_faber_vandereerden} 
is very good up to reduced densities of the order of $0.6$, 
while at higher values the phase of oscillations in the density profile are 
correctly captured, although a slight underestimation of the peak value is observed.
The comparison of our density profiles with those predicted by the ``White Bear'' version of 
the FMT \cite{FMT_WB} shows, as expected, that Rosenfeld's theory gives more accurate estimates
of the oscillation peaks, particularly at high density.
\newline
When the temperature is decreased the contribution of the Yukawa tail to the density profile 
becomes relevant. We compared the density profile obtained within our DFT approximation 
with the MC simulation data for $\zeta\sigma=1.8$ at temperature $T=2$ and for $\zeta\sigma=3$
at $T=1.004$.
Figure \ref{fig:T2_diversirho} shows that the WDA estimate is remarkably accurate 
at reduces densities $0.4$ and $0.5$; small deviations from the MC simulation 
data appear at reduced density $0.7$.  We note that at $\rho_b\sigma^3=0.7$ 
the contact reduced density is overestimated of about $0.2$ with respect to simulation data, 
even if the contact theorem is verified with a relative error of the order of $10^{-5}$. 
This disagreement in the contact value is due to different estimates of the grand canonical 
potential per unit volume of the homogeneous fluid and it is compatible with the the spread 
in the values of the bulk pressure obtained within different simulation 
techniques \cite{MC_thermodynamics_yukawa_99}.
In panel a) of Figure \ref{fig:profili_forze_louis} we compare our results for the density profile of 
the HCY fluid characterized by $\zeta\sigma=3$ with the recent MC simulations from 
Ref. \cite{louis_yukawaMC_plates}. This figure shows that at relatively low densities 
the agreement between our approximation and MC predictions is remarkable also 
for attractive potentials of shorter range.
In particular we predict accurately the kink in the density profile at $\rho_b\sigma^3 = 0.191$, 
which is only qualitatively reproduced within mean field approximation \cite{louis_yukawaMC_plates}.

\begin{figure}
\includegraphics[width=7cm,clip]{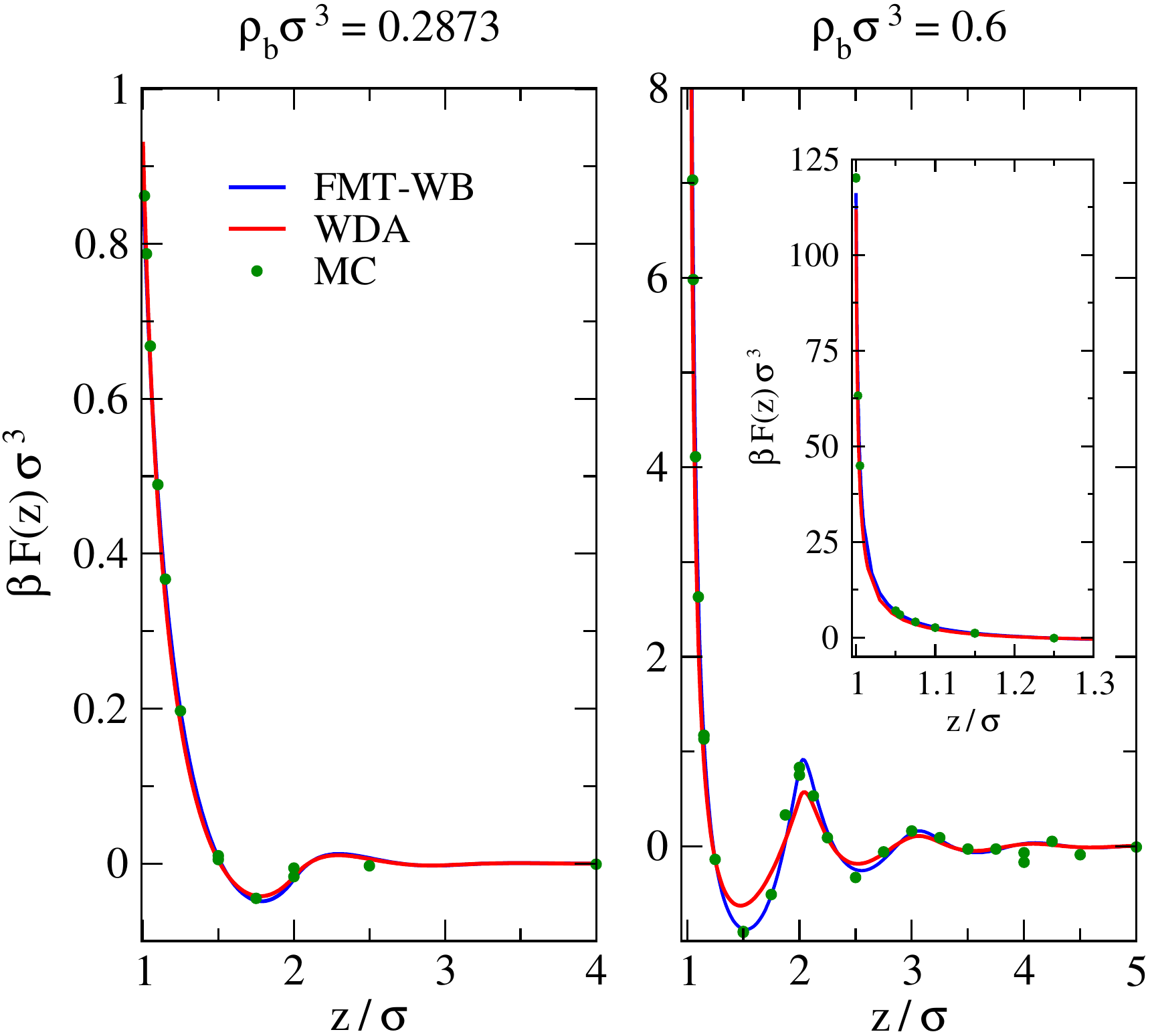}
\caption{Force per unit surface and $k_\mathrm{B}T$ between two infinite planar hard walls 
immersed in a fluid of hard spheres of diameter $\sigma$ for two values of the bulk reduced density. 
The value of the force within the WDA approximation has been obtained via Eq. (\ref{eq:solvation_rho}). 
The FMT-WB result comes from the minimization of the ``White Bear'' version of the FMT 
functional \cite{FMT_WB} via Eq. (\ref{eq:solvation_rho}). The MC data is taken from 
Ref. \cite{wertheim_HW_MD}. The inset highlights the behavior of the force 
at $\rho_b\sigma^3=0.6$ at small distances.}
\label{fig:forza_hs_piani}
\end{figure}

The minimization of the grand canonical functional provides both the value of the contact 
density $\rho^{(L)}_w$ and of the grand free energy. 
It is therefore possible to obtain the depletion force either by calculating the derivative w.r.t. $L$ 
of the grand potential, as in Eq. (\ref{eq:solvation_derivata}), or by making use 
of Eq. (\ref{eq:solvation_rho}). The consistency between the two estimates is a good check for 
the accuracy of the numerical procedure. 
To give an example, for a hard sphere fluid, the relative difference between the two results 
is less than $0.01 \%$ if the absolute value of the force per unit surface and $\mathrm{k}_{\mathrm{B}}T$ 
is larger than $10^{-6}$. Nonetheless at smaller values of the force the result obtained by
differentiation of the grand free energy is less stable, due to errors introduced by the discretization.
\newline
The solvent mediated force acting on two parallel hard walls immersed in a hard sphere fluid, 
is compared with the Monte Carlo data of Wertheim et al. \cite{wertheim_HW_MD} as well as the 
predictions based on FMT in Fig. \ref{fig:forza_hs_piani}, showing a nice agreement also at 
relatively high densities.
At reduced density $0.2873$ the force maximum per unit surface is of the order of 
$\mathrm{k}_{\mathrm{B}}T/\sigma^3$, and the oscillations due to the packing of the hard spheres 
are damped within two or three diameters. Furthermore, at this value of the reduced density, 
a small deviation between WDA and FMT is present only at the first minimum.
At reduced density $0.6$ the force at distances of the order of the hard sphere diameter $\sigma$ 
is hundred times larger than at $\rho_b\sigma^3=0.2873$. Even this feature is well reproduced by both WDA 
and FMT, as can be seen in the inset. The strong oscillating behavior of the MC data 
is captured by WDA with a correct phase, even if the peak values are a little underestimated, 
whereas FMT behaves considerably better.
\newline
The solvent mediated force per unit area between two walls in a HCY fluid has not been extensively 
investigated by numerical simulations. 
In panel b) of Fig. \ref{fig:profili_forze_louis} we show a comparison between our results 
and the Monte Carlo data of Ref. \cite{louis_yukawaMC_plates}
for a HCY fluid of inverse range $\zeta\sigma=3$ at $T=1.004$ and $\rho_b\sigma^3=0.191$.
Even if the net force is quite small, our prediction agrees very well with the numerical simulations
at all values of the wall separation. We stress that, particularly at small distances, the WDA force is much 
more accurate than any mean field perturbation method (see Ref. \cite{louis_yukawaMC_plates}). 

The detailed comparisons of our novel DFT with both numerical simulations and 
state of the art theories allow to conclude that in the high temperature limit
the present WDA is able to correctly reproduce 
the density profile and the effective interactions between the hard walls with a very satisfactory 
accuracy up to reduced densities of about $0.5$. Moreover, this formulation of WDA appears to
be the best available DFT for a HCY fluid at finite temperature. 

\section{Results}
\subsection{Slab geometry}
\begin{figure}
\includegraphics[width=7cm,clip]{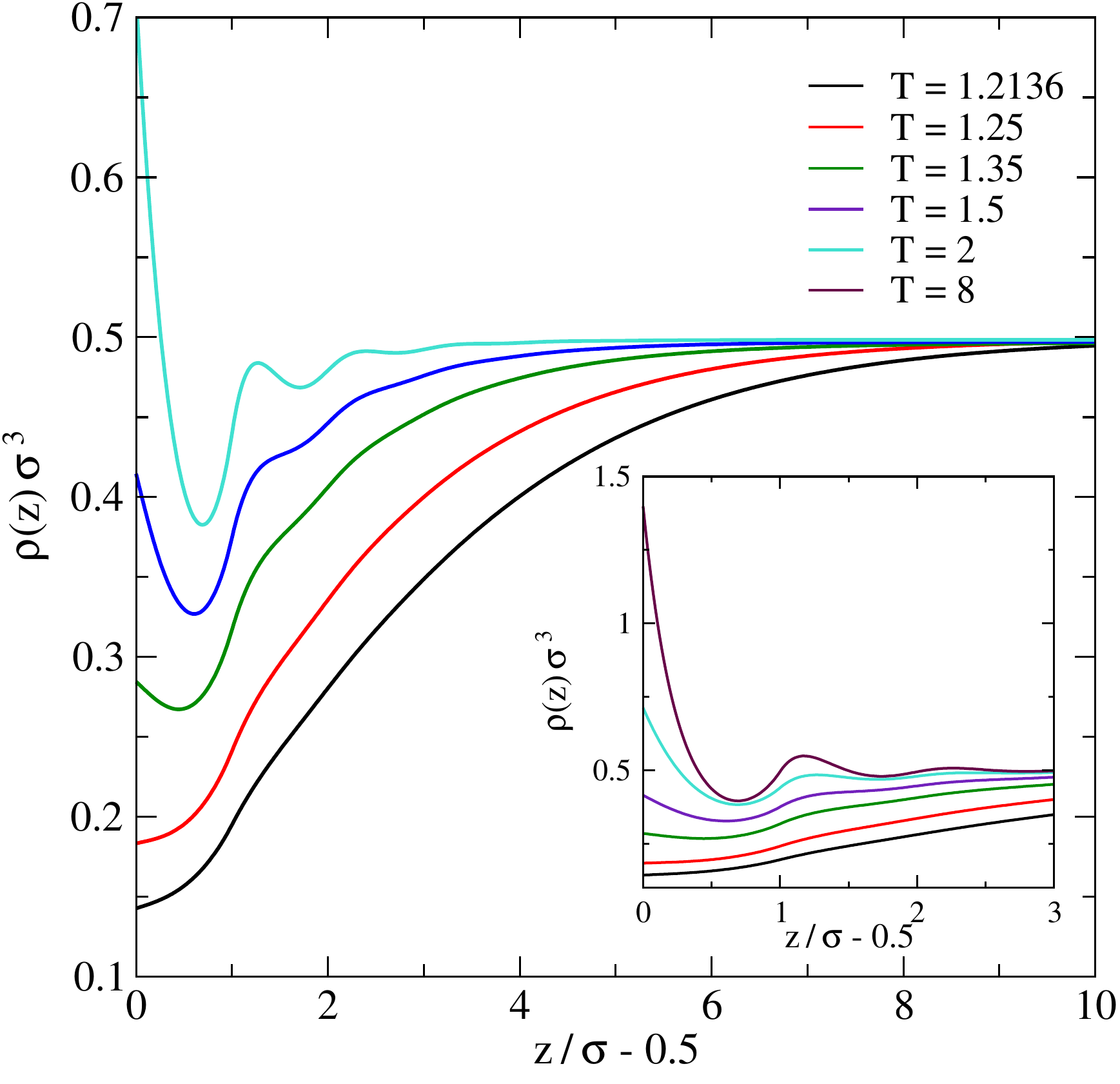}
\caption{Density profile of a HCY fluid ($\zeta \sigma=1.8$) at a single hard wall at bulk 
reduced density $\rho_b{\sigma}^3=0.5$ and different values of the temperature. 
The inset shows the same data, also including the density profile at temperature 
$T=8$ using a different scale.
}
\label{fig:profili_05}
\end{figure}

\begin{figure}
\includegraphics[width=7cm,clip]{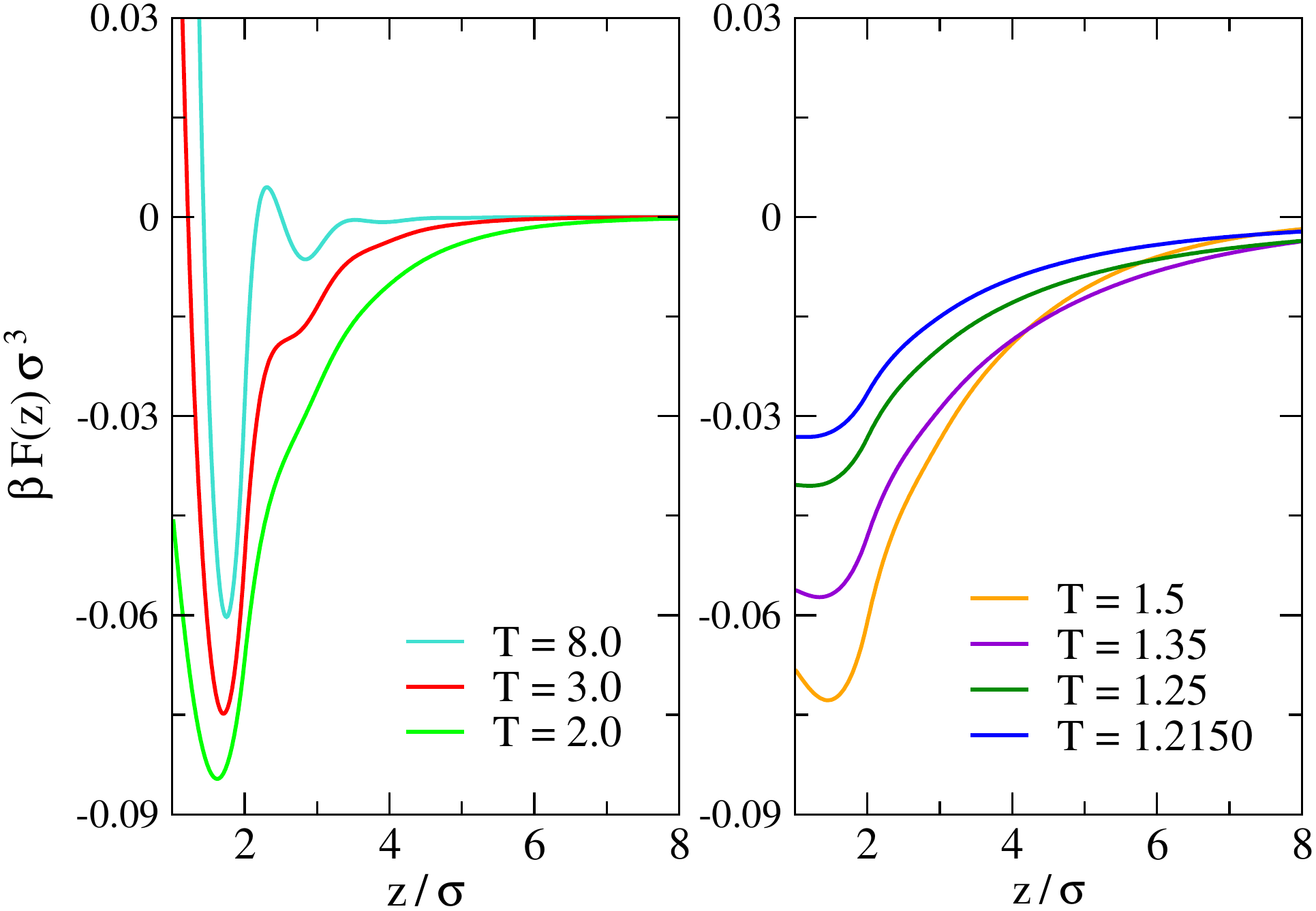}
\caption{Force per unit surface acting between two hard walls immersed in a HCY fluid ($\zeta\sigma=1.8$) 
obtained from the minimization of the WDA functional using Eq. (\ref{eq:solvation_rho})
at $\rho_b\sigma^3=0.3152$ and different values of the temperature.}
\label{fig:forze_rhocritico}
\end{figure}

\begin{figure}
\includegraphics[width=7cm,clip]{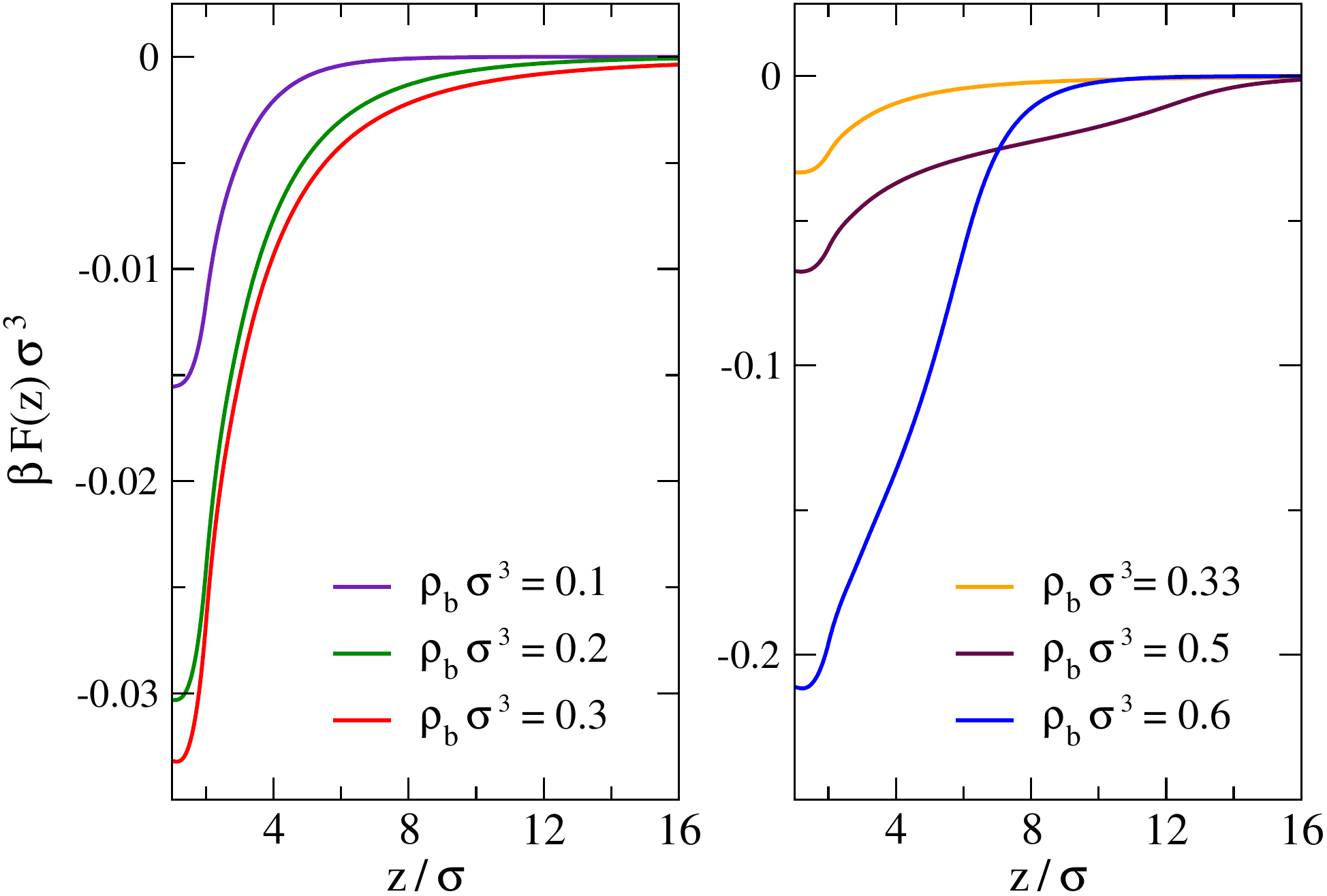}
\caption{Force per unit surface acting between two hard walls immersed in a HCY fluid 
($\zeta\sigma=1.8$) obtained from the minimization of the WDA functional using Eq. (\ref{eq:solvation_rho})
at $T=1.2155$ and different values of the reduced density.}
\label{fig:forze_tcritico}
\end{figure}

We performed the minimization of the WDA density functional at several values of 
temperatures and reduced bulk densities for a HCY fluid of inverse range $\zeta\sigma=1.8$ 
confined between two hard walls. At high temperatures the system behaves like a hard sphere 
fluid, whereas when the temperature is decreased the contribution of the Yukawa tail becomes 
more and more relevant, and the strongly oscillating character of both the density profiles 
and the solvent mediated force is lost.
\newline
Figure \ref{fig:profili_05} shows the dependence of the density profile on temperature at 
fixed bulk reduced density $\rho_b\sigma^3=0.5$. At reduced temperature $T=8$ the system 
behaves like a hard sphere fluid. As the temperature is lowered towards its critical 
value, the density profile gradually becomes monotonic
losing the oscillating features typical of hard spheres 
and the density at contact assumes values four 
times lower than the bulk density. The range of the perturbation produced by the wall extends 
at larger and larger distances as the temperature approaches $T_c$, giving rise to a region 
where a kind of drying of the wall can be observed.
\newline
The attractive tail in the pair interaction of the HCY fluid smoothes the density profile 
reducing the layering of particles. As a consequence, the effective force between the two walls loses
the strongly repulsive peak present at $z\sim \sigma$ 
when the interaction between the fluid particles is purely hard sphere 
(see Fig. \ref{fig:forza_hs_piani}). Fig. \ref{fig:forze_rhocritico} shows the force per unit 
surface between the walls for different values of the temperature 
at the critical bulk reduced density $\rho_c\,\sigma^3 = 0.3152$. 
The repulsive contribution to the interaction force, present at $T=8$ gradually disappears 
at lower temperatures and the force becomes purely attractive and monotonic, confirming the 
findings of the numerical simulations in a different model \cite{gnan_zacca_scio_symmboundaries}.
By approaching the critical temperature ($T_c \sim 1.21353$) the 
effective force becomes weaker and weaker at short distance, as can be seen in the right panel 
of Fig. \ref{fig:forze_rhocritico}: its amplitude reduces almost by a factor two 
due to a $10\%$ change in temperature. However, a closer look to the long distance tail of the 
solvent mediated force shows that its range indeed increases close to the
critical temperatures, as expected 
on the basis of scaling arguments. However this occurs at very large separations ($L>26\sigma$ for the 
data shown in the figure). 
\newline
Fig. \ref{fig:forze_tcritico} shows the force per unit surface between the walls in different 
density regimes when the value of the temperature is close to $T_c$. We note that when the 
bulk density is higher than $\rho_c$ the force is an order of magnitude larger than 
for $\rho_b< \rho_c$. The force is monotonic and purely attractive for all values of 
the reduced density and its range grows near $\rho_c$, as expected.

\begin{figure}
\includegraphics[width=7cm,clip]{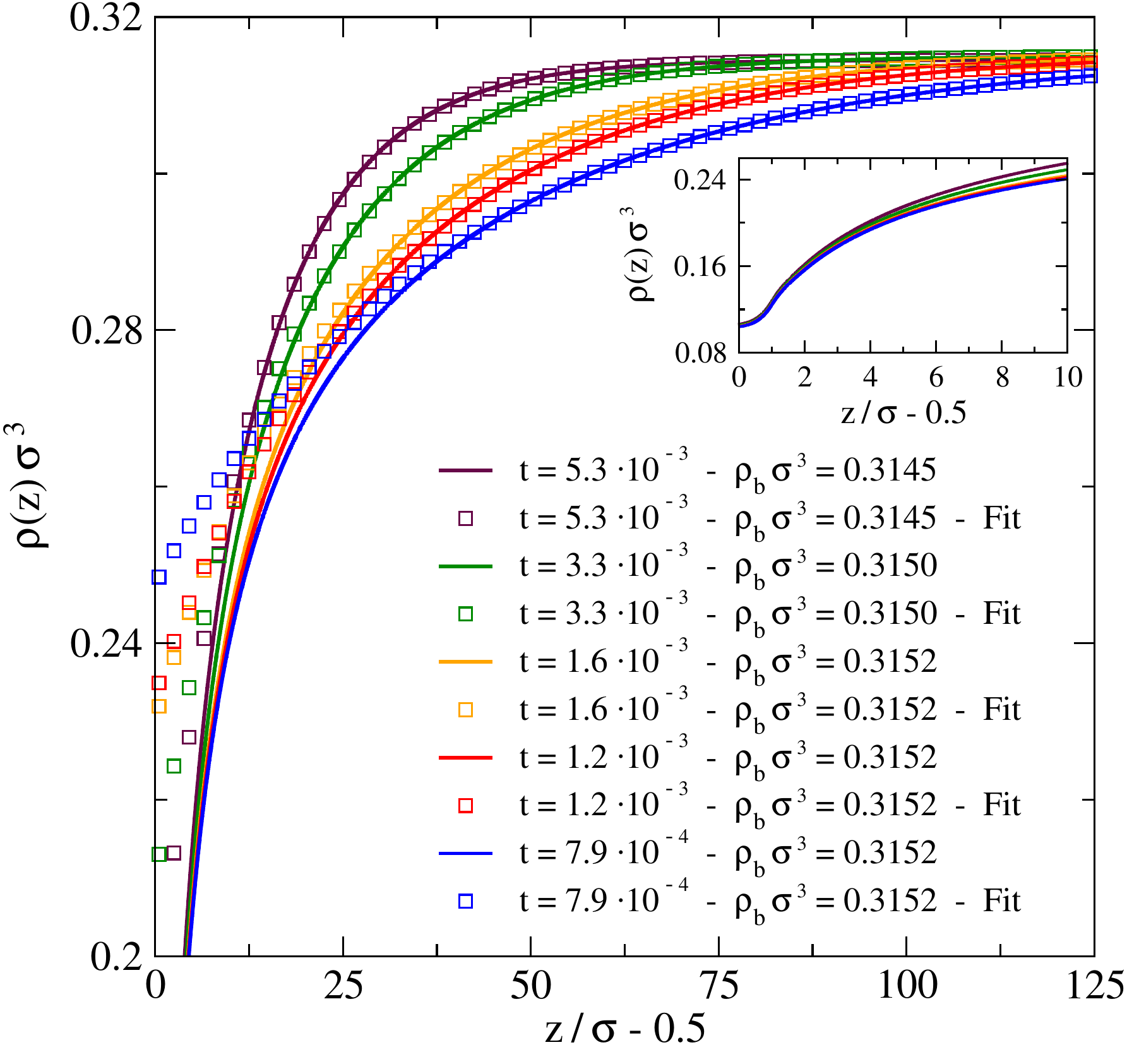}
\caption{Lines: density profile of a HCY fluid ($\zeta \sigma=1.8$) at a single hard wall 
in the critical region. Here $t=(T-T_c)/T_c$. 
Points: fits of the density profiles according to Eq. (\ref{eq:fit_profilo}) 
performed at distances larger than three times the 
correlation length of the homogeneous HCY fluid at the same temperature and bulk density. 
$\rho_b$, $A$ and $\xi$ are free fitting parameters and the results obtained for 
$\rho_b$ and $\xi$ agree well with the bulk values of density and correlation 
length respectively (the accuracy is better than $1\%$ for the bulk density and $5\%$ for the correlation lenght). The bulk correlation lengths obtained from the fitting procedure
are $\xi=41.1\sigma,31.8\sigma,26.5\sigma,17.3\sigma,12.9\sigma$,
from the lowest to the highest reduced temperature. The inset shows a magnification of the same density profiles at 
short distances. 
}
\label{fig:fit_profili_rhocritico}
\end{figure}

\begin{figure}
\includegraphics[width=7cm,clip]{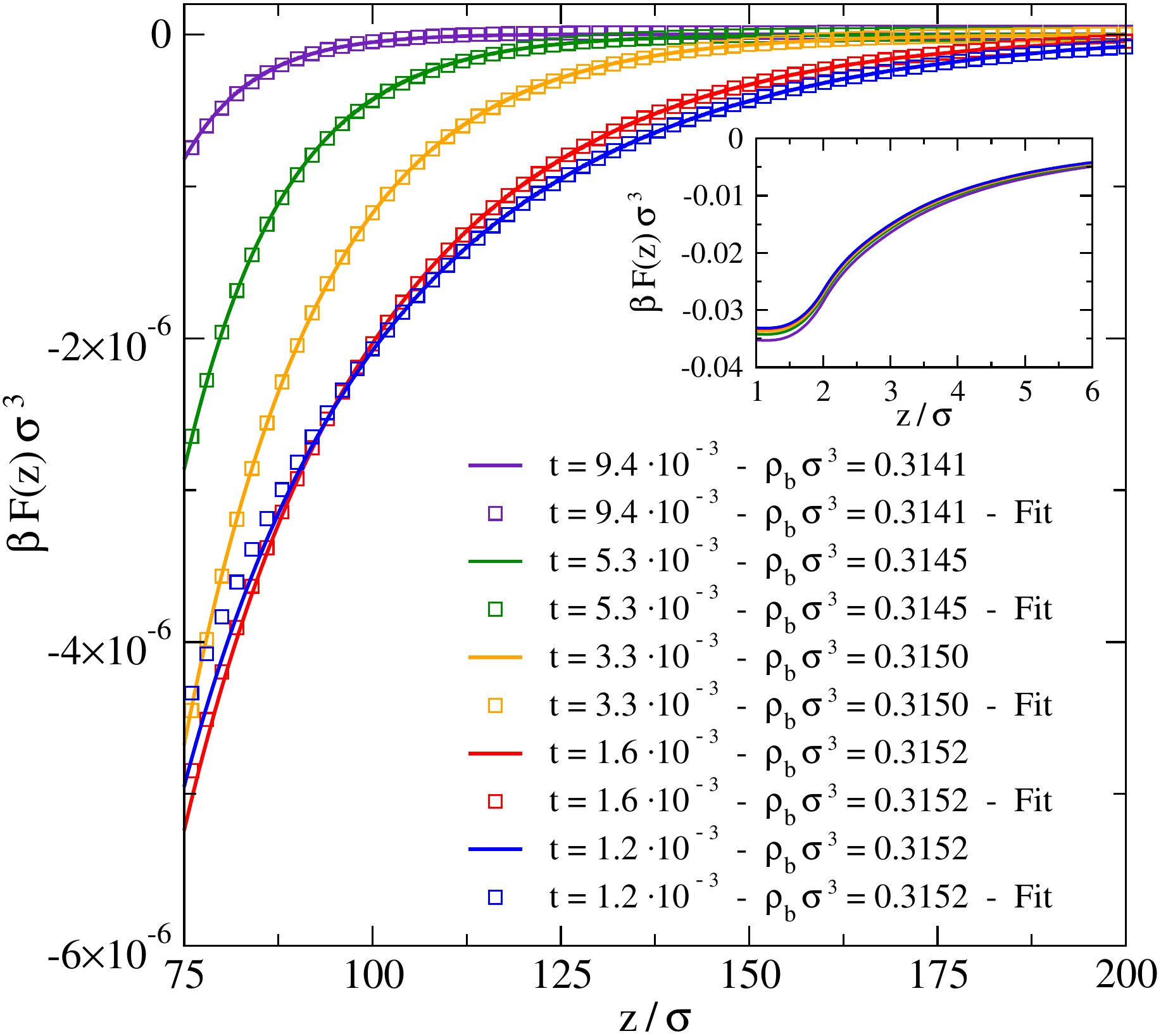}
\caption{
Lines: force per unit surface of a HCY fluid ($\zeta \sigma=1.8$) at a single hard wall
along the line $\tilde{\rho}(T)$. Here $t=(T-T_c)/T_c$.
Points: fits of the force according to Eq. (\ref{eq:fit_forza}) 
performed at distances larger than four times the bulk correlation length of the homogeneous HCY 
fluid at the same temperature and bulk reduced density. 
$f_0$, and $\xi$ are free parameters in the fitting procedure. 
The result obtained for $\xi$ agrees well with the bulk value of the correlation length (the accuracy of the fit is better than $3\%$). 
The bulk correlation lengths obtained from the fitting procedure
are $\xi=31.8\sigma,26.5\sigma,17.3\sigma,12.9\sigma,9.2\sigma$,
from the lowest to the highest reduced temperature. 
The inset shows the force profile at short distance.}
\label{fig:fit_forze_rhocritico}
\end{figure}

In lattice fluid models the coexistence curve is symmetric about the critical temperature 
and the critical isochore coincides with the locus $\rho(T)$ of the maxima of the isothermal 
susceptibility. For such systems, according to the Renormalization Group terminology, the path to the 
critical point orthogonal to the relevant odd
operator coincides with the critical isochore. In a real fluid the coexistence curve 
is asymmetric about the critical isochore. In this case a good approximation for the same path 
is given by the line $\tilde{\rho}(T)$ in the phase diagram defined as 
the locus of the points $(\rho,T)$ such as:
\begin{equation}
\tilde{\rho}(T)=\max_{\rho}\bigl\{ \rho\,{\chi_{T}}\bigr\},
\label{eq:crit_line}
\end{equation}
where $\chi_T$ is the isothermal compressibility. 
\newline
Fig. \ref{fig:fit_profili_rhocritico}
shows the density profile of the HCY fluid at a hard wall 
along the line $\tilde{\rho}(T)$. Its behavior at distances 
larger than the bulk correlation length is well fitted by an exponential of the form:
\begin{equation}
\rho(z)=\rho_b+A\,\mathrm{e}^{-z/\xi},
\label{eq:fit_profilo}
\end{equation}
where $\rho_b$ is the bulk density of the fluid, $A$ is a negative amplitude factor and $\xi$ 
is the bulk correlation length. This exponential decay of the density profile is observed 
also if the system is far from the critical region and is probably related to the location in the bulk phase diagram of the 
point we investigated with respect to the \hbox{Fisher-Widom} line \cite{fisher_widom_WDA,fisher_widom_yukawa}.
\newline
Following the argument of Ref. \cite{fisher_widom_WDA}, we expect that the exponential long range behavior of the density profile 
reflects in an analogous exponential decay of the force between the two walls. 
Provided we do not cross the \hbox{Fisher-Widom} line, this decay should be present both far from the critical point and in the critical
region, where it agrees with the predictions of the theory of the critical Casimir effect \cite{casimir_TIRM_PRE}:
\begin{equation}
\beta {f(z)} = f_0 \,\mathrm{e}^{-z/\xi}.
\label{eq:fit_forza}
\end{equation}
The exponential decay of the solvation force is indeed confirmed
by the exact solution of a \hbox{two-dimensional} \cite{evans_ising2d} 
Ising slab under symmetry breaking boundary conditions and was observed 
in Monte Carlo simulations of three dimensional simple 
fluids \cite{gnan_zacca_scio_symmboundaries,gnan_zacca_scio_allboundaries}.
Fig. \ref{fig:fit_forze_rhocritico} shows the long distance exponential 
decay of the force per unit surface and $\mathrm{k}_{\mathrm{B}}T$ between two planar hard walls
mediated by a HCY fluid along the critical line $\tilde{\rho}(T)$. The force obtained with 
our approach is very well fitted by Eq. (\ref{eq:fit_forza}) at large distances 
($z\gg\xi(\rho_bT)$), 
whereas at short distances, where depletion effects become relevant, the solvent mediated 
force, always attractive, displays a plateau (see inset). 

\subsection{Effective interaction between spherical particles}
The same WDA formalism previously introduced may be generalized to other interesting geometries, in addition
to the planar one. Most importantly, it can be used to evaluate the effective interaction between 
two spherical particles in a solvent, with obvious applications to the study of aggregation in 
colloidal suspensions. 

\begin{figure}
\includegraphics[width=7cm,clip]{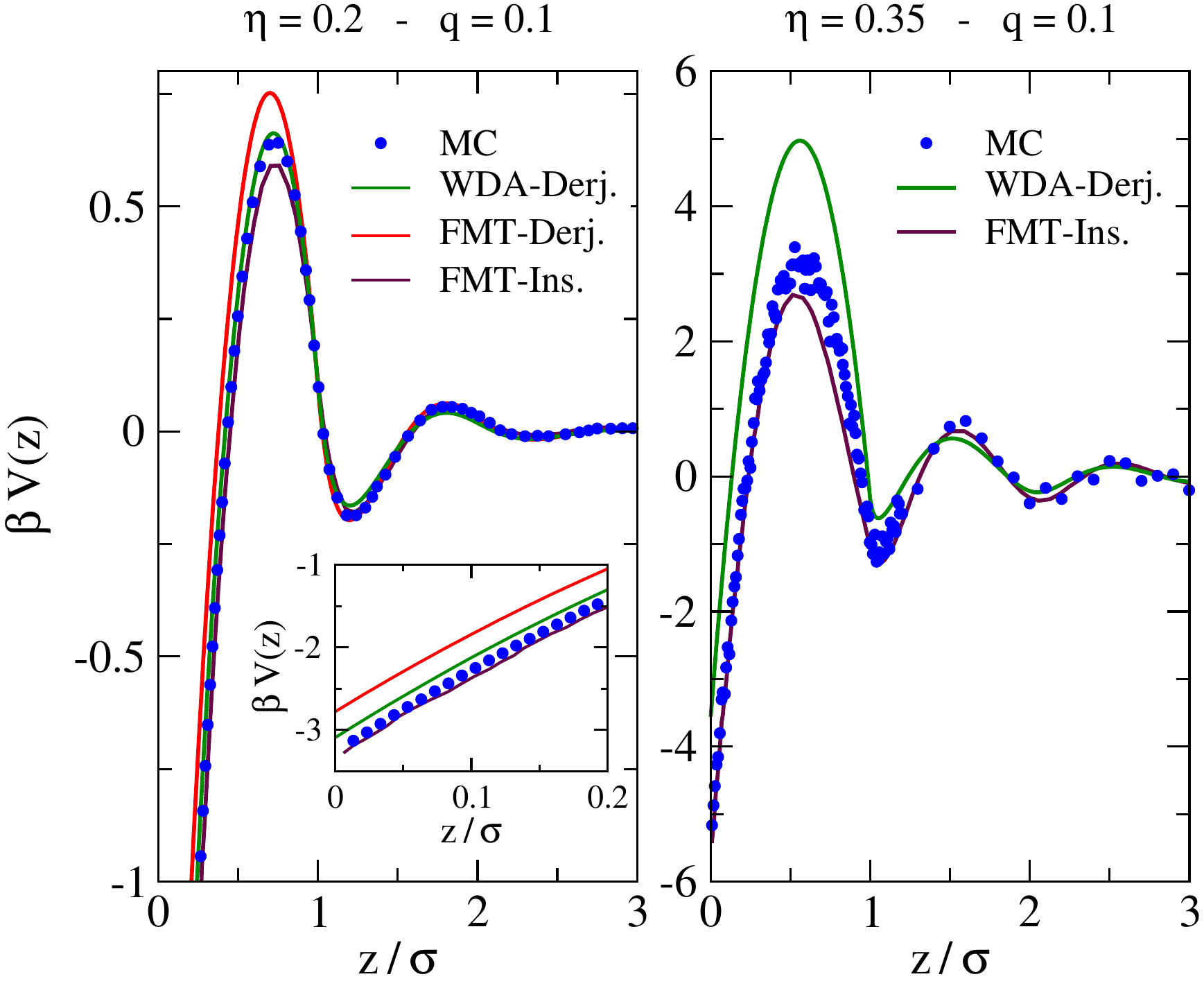}
\caption{Depletion potential per unit $\mathrm{k}_{\mathrm{B}}T$ between two hard spheres in 
a fluid of smaller hard spheres (size ratio $q=0.1$) at bulk packing fraction 
$\eta=\frac{\pi}{6}{\rho}_{b}\,{\sigma}^3$ equal to $0.2$ (left panel) and $0.35$ (right panel).
MC (points) and FMT predictions obtained with the insertion trick (purple lines) 
are taken from Ref. \cite{ashton_wilding_roth_evans_2011}. The red line at $\eta=0.2$ is obtained 
by use of Derjaguin approximation starting from the solvation force between two 
planar hard walls evaluated from the FMT-WB \cite{FMT_WB} approximation. 
The green lines represent the depletion force obtained within Derjaguin approximation 
when the solvation force between the walls is given by the present WDA approximation.}
\label{fig:depletion_sferedure}
\end{figure}

The direct minimization of the DFT, although numerically feasible, represents 
a task considerably more complex than in planar geometry. Therefore, in this first application of the 
formalism, we have chosen to resort to the simple but effective Derjaguin approximation \cite{derjaguin_tedesco_34}, 
which allows to express the interaction between two convex objects 
starting from the knowledge of the force between two planar walls, independently on the 
physical origin of the force.
According to this approximation, the force $F_{\mathrm{D}}$ between two spheres of radius $R$ can be written as:
\begin{equation}
F_{\mathrm{D}}(L)=\pi R\int_{L}^{+\infty}\mathrm{d}z f(z),
\label{eq:boris}
\end{equation}
where $L$ is the minimal surface to surface distance between the sphere and $f(z)$ 
is the force per unit surface between the two walls at distance $z$.
This approximation gives accurate results provided $L\ll R$ and if the 
interaction potential between the two walls decays rapidly at large distances. 
\newline
When the force between the walls is mediated by a hard sphere fluid with particles of 
diameter $\sigma$ it is possible to show that Derjaguin's expression is 
the best approximation of the true depletion interaction without taking in account 
curvature effects \cite{depletion_goz_eva_diet_98} and it is accurate in the 
limit of $q\ll 1$, where $q=\sigma/2R$ is the size ratio.
The limits of Derjaguin approximation when applied to depletion interactions is a debated issue. 
Particularly, it is a matter of discussion the size ratio at which 
Derjaguin's theory starts to fail, and how its accuracy depends on the concentration of depletant. 
In Figure \ref{fig:depletion_sferedure} we compare the prediction for the depletion potential 
$\beta V(L)$ between two big hard spheres in a fluid of smaller hard spheres obtained both 
by Derjaguin approximation and MC simulations at two different values of the bulk density of 
the smaller particles. 
The predictions for the depletion potential obtained by Derjaguin 
approximation are accurate at $q=0.1$ only at values of the packing fraction of the small 
spheres lower than $0.25$. At $\eta=0.35$ Derjaguin approximation overestimates by about 
$2\mathrm{k}_{\mathrm{B}}T$ both the contact value and the first repulsive peak in the potential, 
while the oscillations at larger values of distance are underestimated when compared to MC data. 
The rather poor performance of Derjaguin approximation 
is probably due to the presence of a strong repulsive 
peak in the solvation force between the two walls at $z\sim\sigma$ 
(see Fig. \ref{fig:forza_hs_piani}) which appears to be a peculiarity of the slab geometry. 
We expect that, for smoother inter-wall effective interactions, the agreement would be considerably better. 

The results presented above show that Derjaguin approximation can be safely adopted is the size 
ratio between the depletant and the colloid is sufficiently small and up to moderate 
densities of depletant (i.e. $q<0.1$ and $\rho_b\sigma^3<0.4$), and we expect that similar 
considerations apply when the depletant is a Yukawa hard core fluid. 
Within the limits of validity of the Derjaguin approximation, 
we can determine the solvent mediated potential $v_{\mathrm{eff}}(r)$ between two hard spheres of 
radius $R$ immersed in a YHC fluid in order to examine the phase stability of 
such a colloidal suspension. 
\newline
According to Noro-Frenkel extended law of corresponding states \cite{noro_extended_2000}, 
fluids characterized by short ranged interaction potentials obey the same 
equation of state, when expressed in terms of reduced variables.
In particular, it was observed that the dimensionless second virial coefficient
\begin{equation}
B_{2}^{*}(T)=\frac{B_{2}(T)}{B_{2}^{\mathrm{HS}}}=
1+\frac{3}{8R^3}\int_{2R}^{\infty}\mathrm{d}r\,r^2\left(1-\-\mathrm{e}^{-\beta v_{\mathrm{eff}}(r)}\right),
\end{equation}
where $B_{2}^{\mathrm{HS}}$ is the second virial coefficient of a hard sphere system with particles of radius $R$, assumes a value of about $-1.6$ at the critical point independently on the 
particular form of the interaction and that its value remains constant in a relatively large 
density range across the critical point \cite{critical_second_reduced_virial}. 
It is therefore possible to estimate the the gas-liquid spinodal line for 
a system of hard sphere colloidal particles dispersed in a HCY fluid,
by evaluating their reduced second virial coefficient. 
\begin{figure}
\includegraphics[width=7cm,clip]{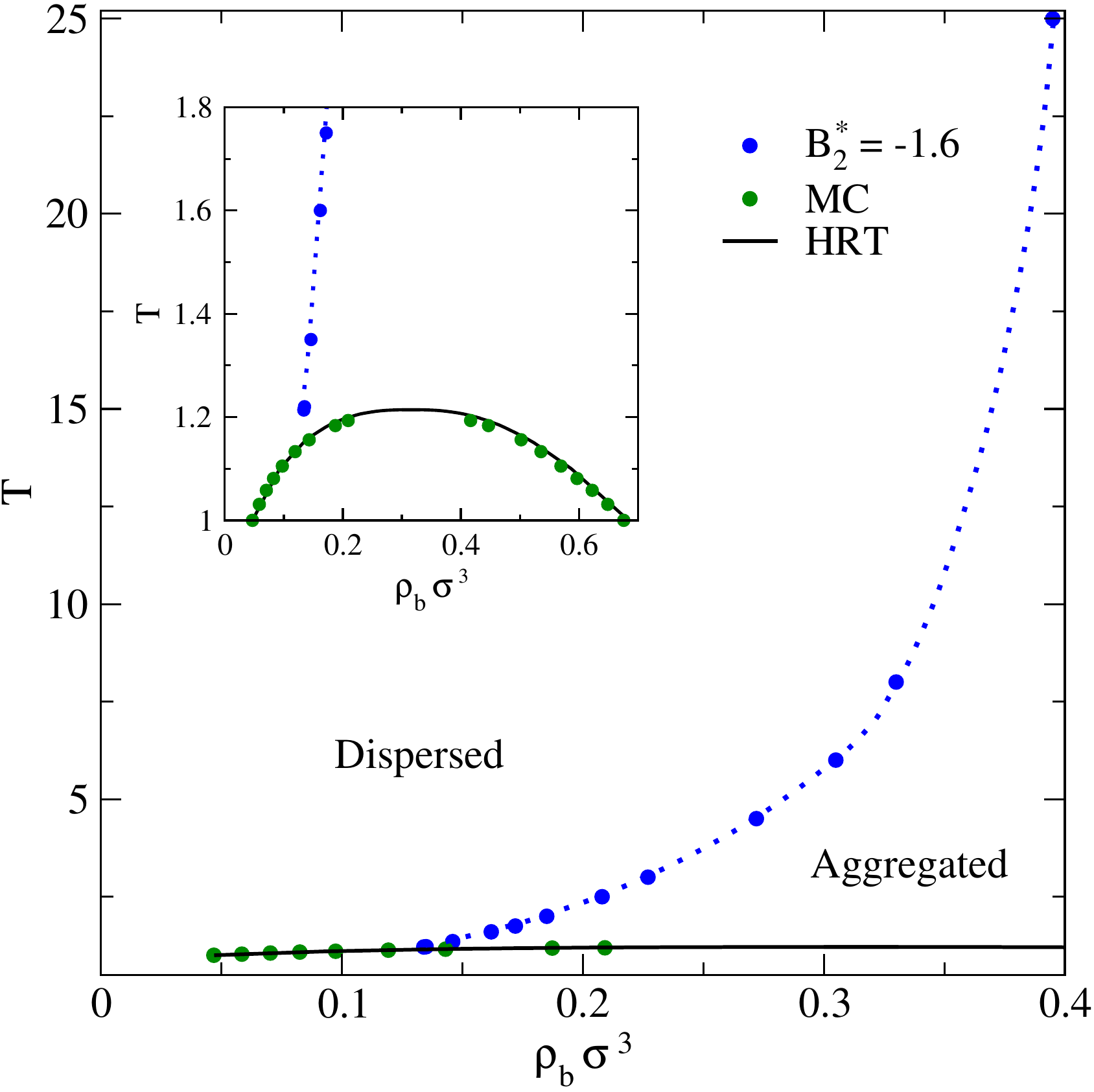}
\caption{Phase diagram of the HCY model ($\zeta\sigma=1.8$) in the $(\rho_bT)$ plane. 
The HRT results for the coexistence curve \cite{smoothcutoff_2008}
are shown by a black line. Green dots represent the MC data from  
Ref. \cite{pini_stell_wilding_datayukawa18}. Blue points show the aggregation boundary of 
two big hard spheres (size ratio $q=\sigma/(2R)=0.1$) predicted on the basis of our 
WDA plus Derjaguin approximation and Noro-Frenkel criterion. 
The dashed line connecting the points is a guide to the eye: aggregation takes place 
on the right of this boundary.  
The inset shows the same phase diagram in a wider density interval.}
\label{fig:aggregation}
\end{figure}
The blue points in Figure \ref{fig:aggregation} identify the phase separation 
line of a HS fluid induced by the presence of a depletant modeled as a HCY fluid. 
The size ratio between the depletant and the guest HS particles is $q=0.1$. 
At high values of the reduced temperature the phase separation occurs at reduced density of 
about $0.4$, as expected in the limit of HS depletant. When the temperature decreases, 
the concentration of depletant needed to induce phase separation decreases monotonically. 
At the depletant critical temperature we observe phase separation at depletant 
concentrations $\rho_b\sigma^3\sim0.15$ much lower than the critical one. This implies that, 
at this value of the size ratio $q$, the 
phase separation is not related to the presence of long range tails in the effective force,
which characterizes the critical region of the solvent, but is still mainly due to the short range 
attraction generated by the depletion mechanism.
\newline
We note that whenever a direct short range repulsion
is present between the colloidal particles, as for the case of charged systems, the strong  
attraction due to depletion is severely weakened and particle aggregation takes place at considerably
larger solvent densities. Instead, in the critical region, the long range tails of the solvent 
mediated (Casimir) force is not effectively contrasted by the additional short range repulsion.
In extreme circumstances 
(i.e. when the direct repulsion between particles is sufficiently strong),
ordinary depletion may be fully screened and phase separation inhibited. 
However, aggregation is generally expected in a small pocket within the critical region,
due to the emergence of long range Casimir forces. For repulsive wall-solvent interactions, this pocket will 
be centered at solvent densities larger than the critical one, due to the strong asymmetry of the critical 
Casimir forces (see e.g. Figs. \ref{fig:forze_tcritico} and \ref{fig:scala_micro_tcritica}). 

\section{Critical Casimir effect}
\subsection{Force profiles}
The aim of this section is to evaluate the solvent mediated interaction 
induced between two walls when the depletant is in the critical regime.
Thermal fluctuations in fluids occur on a characteristic range, determined by the correlation length $\xi$, 
which usually is comparable with the molecular diameter or with the range 
of the interactions. When approaching a second order phase transition, the range of the fluctuations of the 
order parameter (the particle density in the case of a simple fluid) grows larger 
up to diverging at the critical point.
In this regime the effective force between two bodies immersed in the critical fluid acquires a universal 
form and obeys scaling laws, as many physical properties near criticality do.  
\newline
In 1978 \cite{fisher_wall_1978} Fisher and de Gennes first recognized that a confinement 
of the critical fluctuations of the order parameter gives rise to an universal, long ranged 
\hbox{fluctuation-induced} interaction which they named critical Casimir force.
According to the \hbox{finite-size} scaling approach, 
the universal contribution to the force per unit surface $F_\mathrm{C}$
acting between two infinite plates confining a critical fluid can be 
written as \cite{book_FSS,domb1983phase_barber}:
\begin{equation}
\frac{F_{\mathrm{C}}(t,h;L)}{\mathrm{k}_{\mathrm{B}}T}=\frac{1}{L^3}\Theta\left(\pm s,\pm y\right),
\label{eq:casimir_film}
\end{equation}
where $L$ is the distance between the two walls. The upper sign refers to the 
\hbox{super-critical} temperature (while the lower to the \hbox{sub-critical} one) and the scaling variables ($s,y$)
\begin{equation}
s\equiv \frac{L}{\xi};\qquad \qquad
y\equiv a h|t|^{-\beta\delta}
\end{equation}
are defined in terms of the two scaling fields
\begin{equation}
t=\frac{T-T_c}{T_c}; \qquad \qquad h={\mu-\mu_c}.
\end{equation}
Here $\xi\sim\xi_0^{\pm} t^{-\nu}$ is the bulk correlation length at $h=0$, $a$ is a non-universal 
metric factor and $\nu$, $\beta$ and $\delta$ are the usual critical exponents.
The function $\Theta(\cdot,\cdot)$ is usually referred to as the 
scaling function of the critical Casimir force in planar geometry. 
This function is universal in sense that it depends only on the bulk 
universality class, on the boundary conditions imposed at the confining surfaces and on the 
geometry of the system (which in this case is $\infty^{2}\times L$).
We remark that there is no extra metric factor associated with $L/\xi$ and that 
there is a dependence on the sign of the field $h$ because boundary conditions at the 
walls break the bulk symmetry $h\to-h$.
According to the theory of \hbox{finite-size} scaling, Eq. (\ref{eq:casimir_film}) 
represents the asymptotic decay of the solvent mediated force as $t,h\to 0$ and for $L,\xi \to\infty$.
\newline
The bulk Yukawa fluid under investigation belongs to the 3D Ising universality 
class and the boundary conditions are determined by the affinity of the wall surfaces 
with the fluid particles: if the contact density is less than the bulk density the boundary 
condition is of type $-$, otherwise of type $+$. In this work we only deal with \hbox{super-critical} 
temperatures ($t>0$) and with symmetric $(-,-)$ boundary conditions, which arise for purely 
repulsive interactions between the fluid particles and two identical confining hard walls. 

The analysis of the critical Casimir force and of the related universal scaling function 
$\Theta$ is a rather difficult task both experimentally,
for the small forces involved, and theoretically, for the 
lack of an accurate description of critical fluids in confined geometries. 
Most of the results present in the literature deal with the temperature dependence 
of the Casimir fluctuation induced interaction at zero magnetic field. 
An indirect estimate of the scaling function for the film geometry and 
the 3D Ising universality class, at $h=0$ under $(+,-)$ 
and $(+,+)$ boundary conditions was given in 
Ref. \cite{second_casimir_exp_mixture,scaling_casimir_ising_Rafai} monitoring the 
thickness of a binary fluid film at different temperatures near $T_c$. 
The first direct evaluation of the critical Casimir force was performed in 
2008 \cite{hertlein_direct_2008} for a system consisting of a colloidal particle close to a wall 
immersed in a binary mixture (\hbox{sphere-plate} geometry) of water and lutidine at 
different compositions and for both the relevant boundary conditions. 
This experiment, however, allowed to probe only the 
exponential tail of the scaling function. 
A MC study of the solvent mediated potential between 
two spherical particles in a simple fluid along the critical isochore has been 
performed in Refs. \cite{gnan_zacca_scio_symmboundaries,gnan_zacca_scio_allboundaries} 
with different boundary conditions. However, the determination of the full Casimir 
scaling function could not be obtained in the temperature range examined in the simulations. 
Along the symmetry line ($h=0$), more precise estimates of the 
universal Casimir scaling functions for the 3D Ising universality class and 
film geometry have been obtained
via MC simulations of the Ising model \cite{vasilyev_monte_2007,vasilyev_universal_2009,vasilyev_2011}.
Few theoretical approaches were devised to address this problem: in addition to 
the mean field results \cite{krech_meanfield}, its is worthwhile mentioning 
the extended \hbox{de Gennes-Fisher} local functional method \cite{borjan_sym,borjan_asym} and a 
long wavelength analysis of density functional theory \cite{buzzaccaro_critical_2010,piazza_critical}. 
The latter investigations have been also extended away from the symmetry line ($h \ne 0$) providing 
predictions on the shape of the critical Casimir scaling function in the \hbox{off-critical} case \cite{casimir_meanfield_sphericalcolloids,biondi,piazza_critical}. 
Monte Carlo simulations at $h\neq0$ were recently performed in Ref. \cite{vasilyev_scaling_offcritical}.

\begin{figure}
\includegraphics[width=7cm,clip]{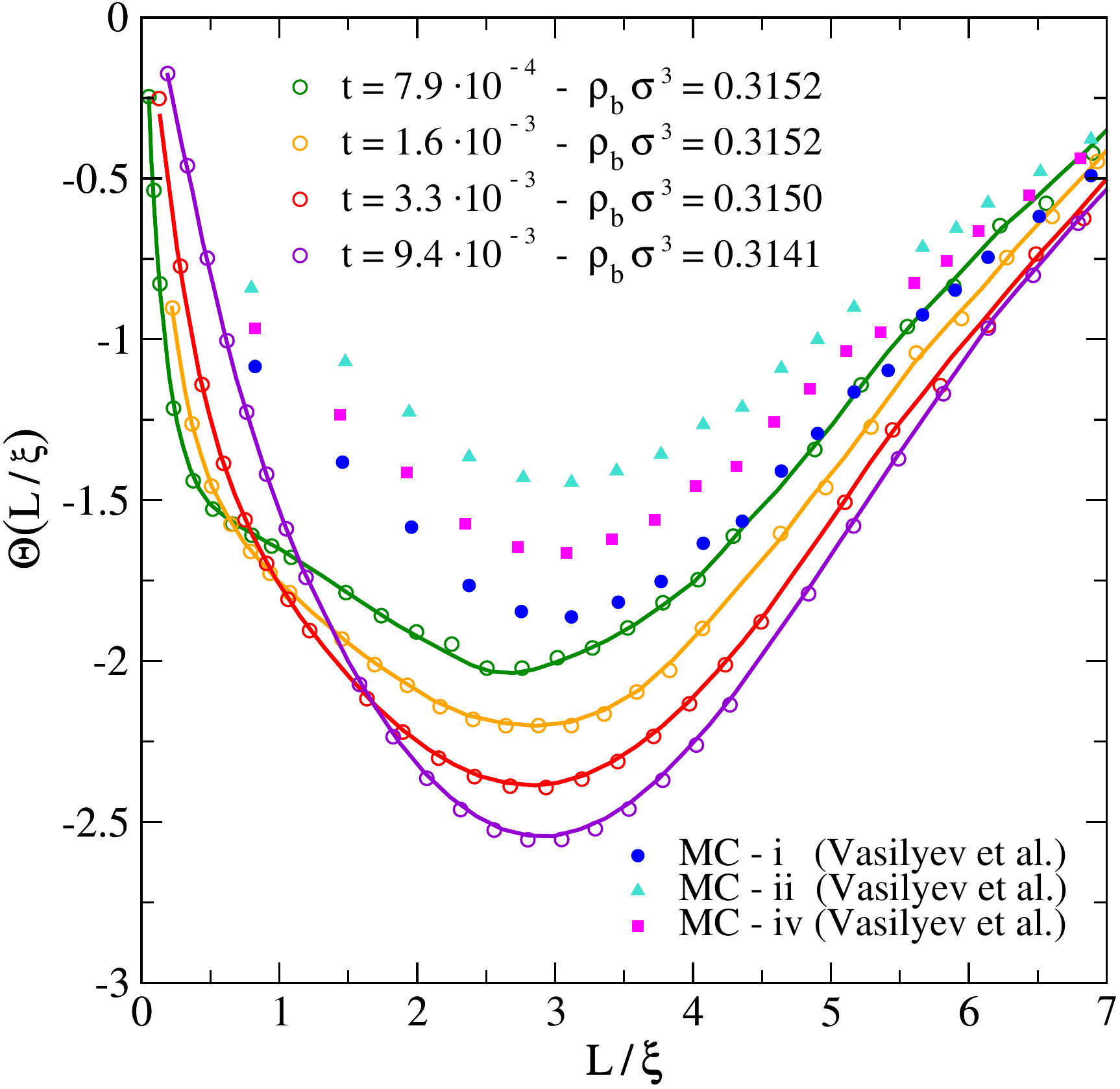}
\caption{
\hbox{Finite-size} estimates of the Casimir scaling function at values of temperatures and density 
along the critical line defined in Eq. (\ref{eq:crit_line}) from the microscopic force 
obtained within the present WDA approximation. The corresponding bulk correlation lengths 
are $\xi=41.1\sigma,26.5\sigma,17.3\sigma,9.2\sigma$, 
from the lowest to the highest reduced temperature.
The lines connecting the points are a guide to the eye.
The MC data are taken from Ref. \cite{vasilyev_universal_2009} and refer to $(-,-)$ 
boundary conditions and the different sets correspond to different estimates of the 
corrections to scaling.}
\label{fig:scala_micro_rhocritica}
\end{figure}

\begin{figure}
\includegraphics[width=7cm,clip]{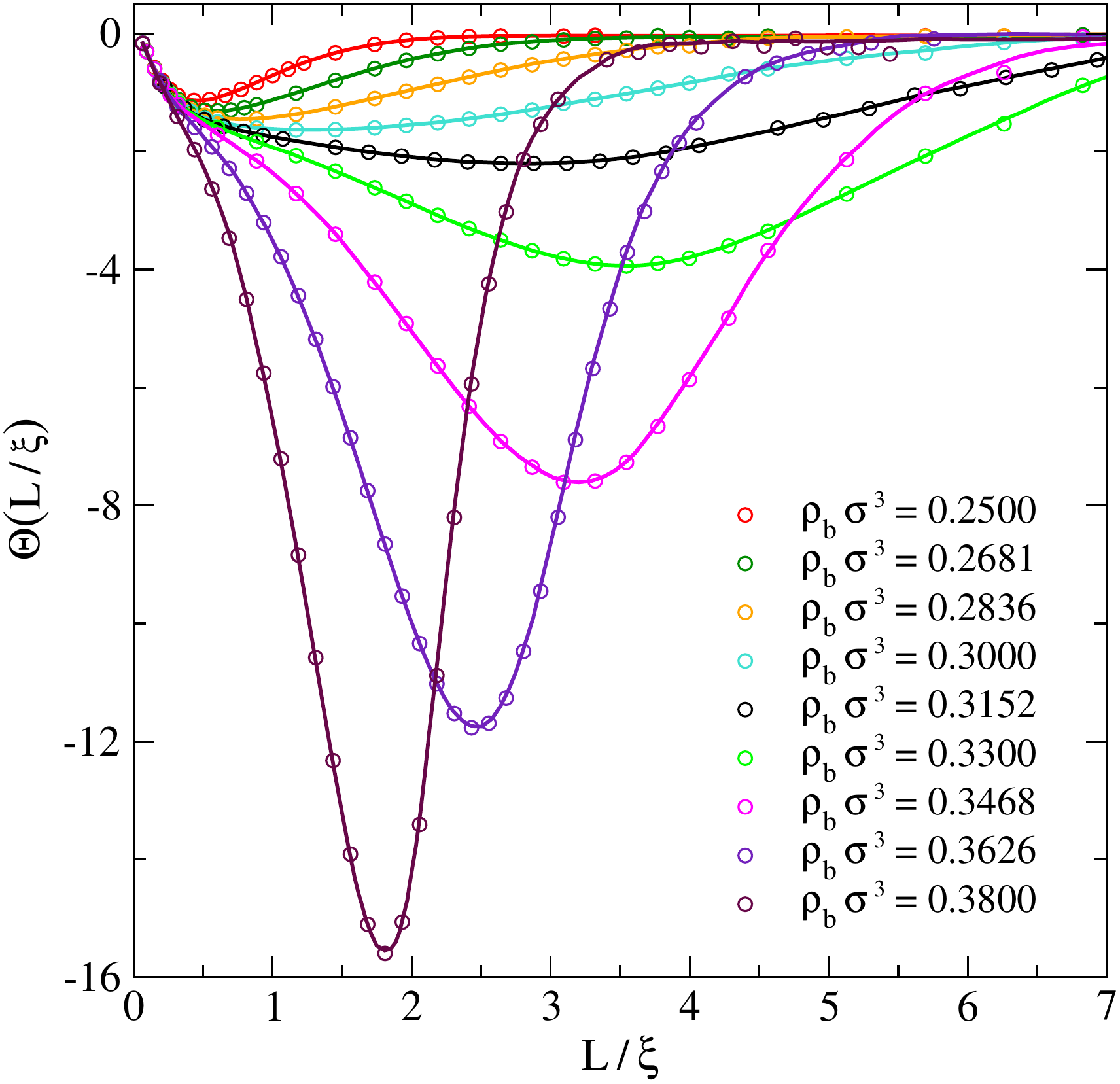}
\caption{Scaling function for the critical Casimir force at $t=1.6\cdot10^{-3}$ and for 
different values of the scaling variable $y$. The lines connecting the points ar a guide to the eye. 
The largest correlation length, $\xi=26.5\sigma$, corresponds to the critical 
reduced density $\rho_b\sigma^3=0.3152$.}
\label{fig:scala_micro_tcritica}
\end{figure}

The WDA approach developed above allows the study of 
this problem starting from the microscopic HCY fluid model
confined between two walls. According to the scaling hypothesis, the effective force per unit surface 
between the two walls $F_\mathrm{C}$ should depend on the physical control parameters $T,\mu,L$ only through 
the combination (\ref{eq:casimir_film}), implying the collapse of different data sets onto the 
same universal curve. 
Figure \ref{fig:scala_micro_rhocritica} shows the scaling function obtained from 
independent calculations at different 
temperatures along the previously defined critical line $\tilde{\rho}(t)$ 
(i.e. $y=0$). Note that, even at reduced temperature $t=(T-T_c)/T_c$ as low as
 $10^{-3}$, our estimates show a marked temperature dependence,
and the data along different isotherms do not collapse as we expected. 
At the lowest temperature we investigated, a significant difference between our prediction and the 
MC simulations of Ref. \cite{vasilyev_universal_2009} suggests the presence of strong corrections to 
scaling. We also remark that the curves at the lowest temperatures develop a kink at small 
values of $L/\xi$, due to the singular behavior of the scaling function at $L/\xi=0$. In fact,
at any given reduced temperature $t\ne 0$, the quantity 
$L^3\,F_\mathrm{C}$ tends to zero as $L/\xi\to 0$, forcing the \hbox{finite-size} estimate of the scaling function to vanish. 
Figure \ref{fig:scala_micro_tcritica} shows the scaling function at fixed temperature near 
$T_c$ for different values of the scaling variable $y= a h|t|^{-\beta\delta}$, corresponding 
to different bulk reduced densities. 
The scaling function is always negative and a strong asymmetry is evident between the curves 
at density above and below $\rho_c$. For positive values of the scaling field $h$ (i.e. $\rho>\rho_c$ 
in our case), the magnitude of the force becomes larger and larger and the peak is 
shifted towards small values of $L/\xi$.

\subsection{Long wavelength analysis}
\begin{figure}
\includegraphics[width=7cm,clip]{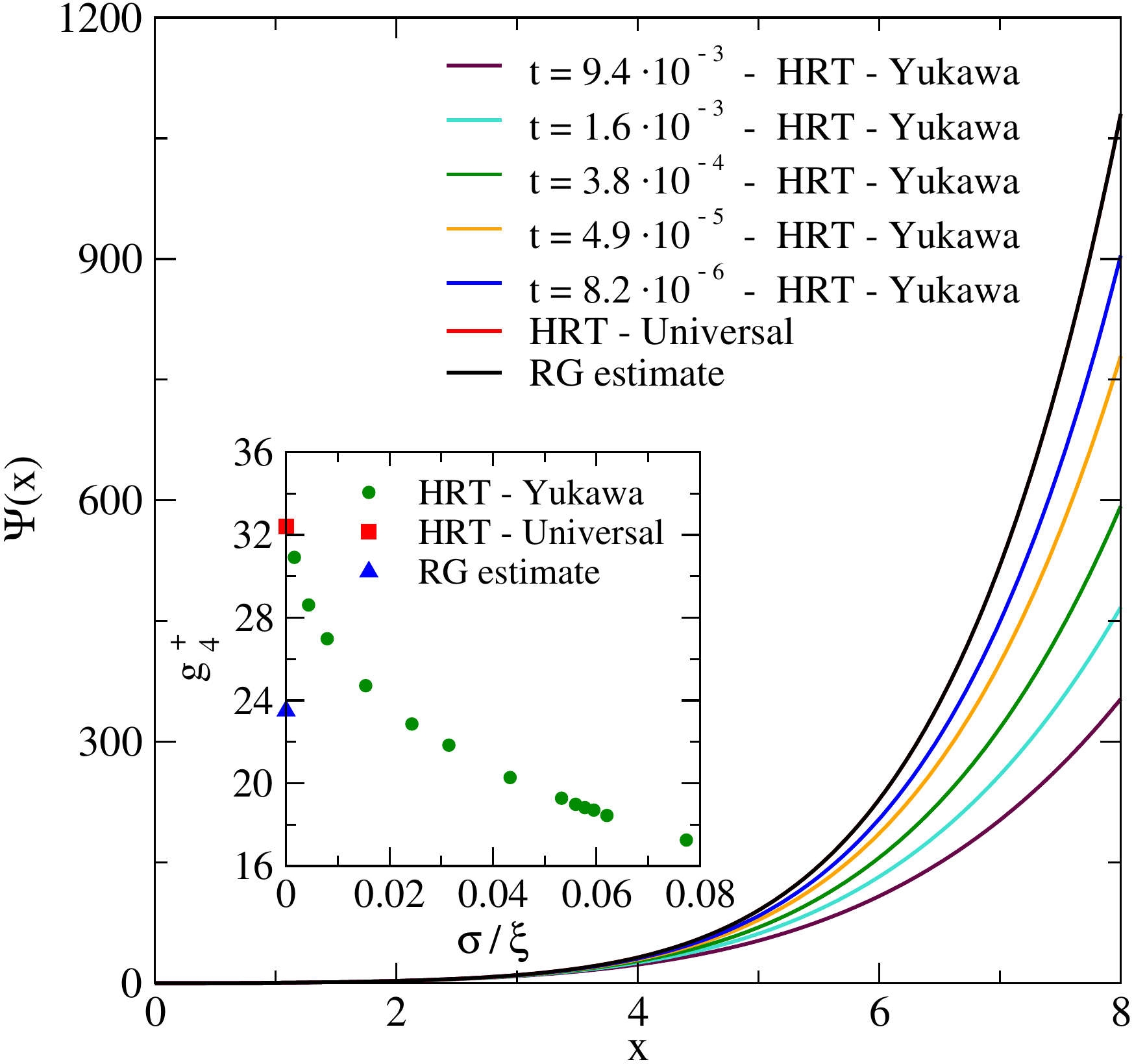}
\caption{Scaling function for the free energy $\Psi(x)$ as predicted by HRT compared with
a parameterization of the exact result from Ref. \cite{pelissetto_critical_2002} (black curve). The red line, showing the
HRT asymptotic result, is identical to the exact result on the scale of the figure. A few rescaled free energies obtained
from the integration of the HRT equations at different reduced temperatures are also shown.
Inset: universal amplitude ratio $g_4^+$ for a HCY fluid at the critical density as a function of the reduced inverse 
correlation length (green points). The usually accepted value \cite{pelissetto_critical_2002} is shown by a blue triangle whereas the 
red square represents the asymptotic HRT value.
}
\label{fig:scalapsi_g4}
\end{figure}

Although the direct numerical evaluation of the critical Casimir scaling function predicted by this
class of DFT is not conclusive, due to severe corrections to scaling, 
an accurate estimate of the asymptotic 
behavior can be obtained by a long wavelength (LW) analysis of the DFT equations. 
In fact, following Ref. \cite{piazza_critical} we note that:
\begin{itemize}
\item the density profile $\rho(z)$ displays reflection symmetry about $z=\frac{L}{2}$, limiting the 
range of interesto to $z\in \left[0,\frac{L}{2}\right]$;
\item when the walls are far apart ($L\gg \sigma$),
the difference between the density profile corresponding to a
wall to wall distance $L$ and its single wall limit, reached for $L=\infty$, is significant only for $z\sim \frac{L}{2}$;
\item as a consequence, the effective force per unit surface $F_{\mathrm{C}}$ (hence the Casimir scaling function)
just depends on the long distance tail of the density profile, which is expected to be a slowly varying
function of the coordinate $z$.
\end{itemize}
Therefore our WDA intrinsic free energy functional can be approximated 
by keeping only the lowest term in a gradient expansion about the bulk density $\rho_b$: 
\begin{eqnarray}
&& \frac{\beta \mathcal{F}\left[\rho_b+\delta n(z)\right]}{\Sigma}=L\,\varphi(\rho_b) + \nonumber \\
&& \,\int {\rm d} z \, \Bigg [ \frac{b}{2} \left ( \frac{{\rm d}\delta n(z)}{ {\rm d} z}\right)^2 + 
\varphi(\rho_b+\delta n(z)) -\varphi(\rho_b) \Bigg ], 
\label{eq:longw}
\end{eqnarray}
where $\varphi(\rho)$ is the free energy density in the bulk, times $\beta$. 
This expression coincides with the long wavelength limit of our WDA functional, the 
stiffness $b$ being related to the range of the direct
correlation function in the homogeneous system $c(r,\rho_b)$:
\begin{equation}
-\int {\rm d} \bm{r}\,c(r,\rho_b) \,\mathrm{e}^{\mathrm{i}\bm{q}\cdot \bm{r}} \longrightarrow 
\frac{\partial^2 \varphi(\rho_b)}{\partial \rho_b^2} + bq^2 +O\left(q^4\right).
\end{equation}
In the presence of short range interactions, the direct correlation function is analytic in $q^2$ away from
the critical point, where it displays a $q^{2-\eta}$ singularity. 
However, within our approximate closure of the 
HRT equations, the critical exponent $\eta=0$ and analyticity is preserved 
also at criticality \cite{smoothcutoff_2008}, keeping the
stiffness $b$ finite in the whole phase diagram. 
This implies that the long wavelength limit of the structure factor of the homogeneous fluid 
follows the Orstein-Zernike ansatz:
\begin{equation}
S(q) \sim \frac{S(0)}{1+\xi^2 q^2} 
\end{equation}
with 
\begin{equation}
\rho_b\,S(0)=\left  [\frac{\partial^2 \varphi(\rho)}{\partial \rho^2}\right ]^{-1} 
\end{equation}
and $\xi^2 = \rho\,S(0)\,b$. Close to the critical point, the HRT bulk free energy density $\varphi(\rho)$ acquires a scaling form:
\begin{equation}
\varphi(\rho_c+\delta \rho) - \varphi(\rho_c)-\beta \mu(\rho_c)\delta\rho 
=t^{d\nu}\,a_{11}\Psi\left(b_1\,\delta \rho\, t^{-\beta}\right),
\label{scaling}
\end{equation}
where $\rho_c$ is the critical density and $\mu(\rho)$ is the chemical potential (the temperature dependence of 
these quantities is understood), while  $a_{11}$ and $b_1$ are non universal
metric factors and $t=(T-T_c)/T_c$ is the reduced temperature. In the following it will be convenient to express the
universal quantities in terms of the scaling field $x=b_1\,\delta \rho\, t^{-\beta}$ instead of the previously defined variable $y$. 
Within our HRT closure, the critical exponents are $\delta=5$, $\beta=0.332$, 
$\nu=0.664$ in $d=3$, which agree within 
$10\%$ with the accepted values. 
The metric factors appearing in the scaling function are 
implicitly defined by the requirement that $\Psi(x)$
has the following expansion at small $x$ \cite{pelissetto_critical_2002}:
\begin{equation}
\Psi(x) \longrightarrow \frac{x^2}{2!} + \frac{x^4}{4!} + O\left(x^6\right).
\label{asinto}
\end{equation}
In Fig. \ref{fig:scalapsi_g4} the asymptotic 
HRT scaling function $\Psi(x)$ is shown together with a parameterization of the exact result for the 3D Ising universality 
class. Although the two curves are indistinguishable on this scale, calculations 
at different reduced temperatures, also shown, suggest the presence of important corrections to scaling. 

The minimization of the long wavelength functional (\ref{eq:longw}) in slab geometry gives rise to a differential equation whose solution 
allows to evaluate the asymptotic decay of the effective force between two hard walls in a critical fluid. The derivation, 
already detailed in Ref. \cite{piazza_critical} and not repeated here, provides a closed form for the critical Casimir scaling function 
in terms of two universal quantities: the bulk free energy scaling function $\Psi(x)$ and the universal amplitude ratio $g_4^+$.
Defining the auxiliary quantity $\sigma(s,x)$ by the implicit relations:
\begin{eqnarray}
\label{sigma1}
&&\sigma(s,x) = -\Psi(x+u_0) + \Psi(x)+u_0\,\Psi^\prime(x);\\
&& s= \int_{u_0}^\infty \frac{\sqrt{2}\, \mathrm{d}u}{\sqrt{\sigma(s,x) +\Psi(x+u)-\Psi(x)-u \,\Psi^\prime(x)}},
\label{sigma2}
\end{eqnarray}
the critical Casimir scaling function in three dimensions is given by:
\begin{equation}
\Theta\left (s,x\right ) = \frac{s^3}{g_4^+} \, \sigma\left (s,x\right ) .
\label{eq:scalingc}
\end{equation} 
The universal amplitude ratio $g_4^+$ is expressed in terms of the non universal 
metric factors previously introduced as
\begin{equation}
g_4^+ = b_1^3\,\sqrt{a_{11}b^{-3}}.
\label{g4}
\end{equation}
Again, the evaluation of $g_4^+$ from the HRT equations displays severe correction to scaling in a HCY fluid, as 
shown in the inset of Fig. \ref{fig:scalapsi_g4}. More importantly, the usually quoted ``exact'' value \cite{pelissetto_critical_2002} 
$g_4^+\sim 23.6$ turns out to differ significantly from the HRT prediction $g_4^+\sim 32.4$. 

\begin{figure}
\includegraphics[width=7cm,clip]{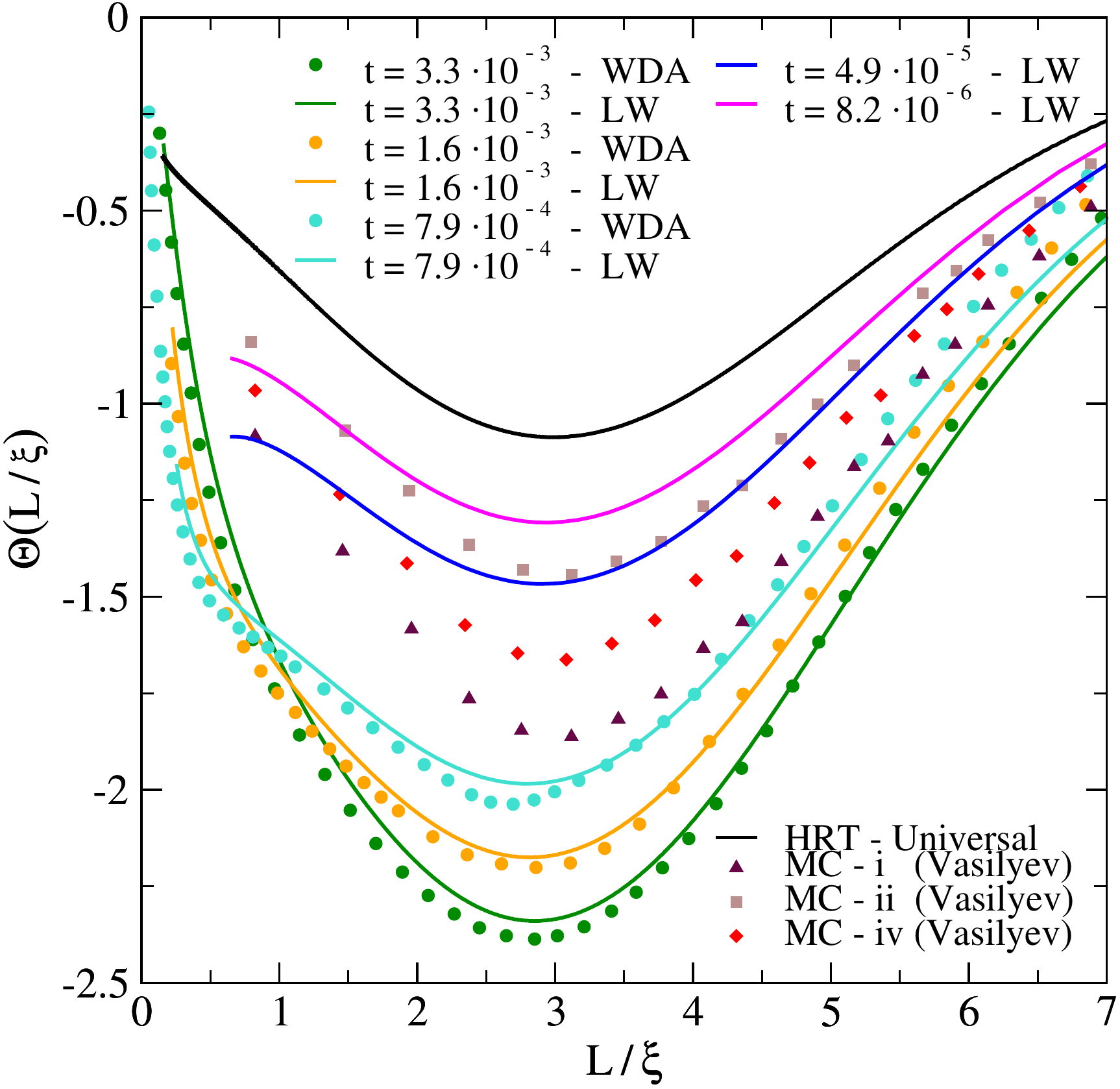}
\caption{Critical Casimir scaling function evaluated at different reduced temperatures. 
Points: results from the 
direct minimization of the microscopic WDA functional. 
Lines: results from the long wavelength analysis, starting from
the universal quantities $\Psi(x)$ and $g_4^+$ evaluated at the same reduced 
temperature as the DFT calculation.  
Black line: asymptotic limit of the critical Casimir scaling function. Points: 
prediction of $\Theta$ from Monte Carlo simulations \cite{vasilyev_universal_2009}.
}
\label{fig:micro_vs_long}
\end{figure}

The asymptotic study of the DFT equations allows to extract the critical Casimir scaling function just from 
bulk quantities via Eqs. (\ref{sigma1}-\ref{eq:scalingc}). It is then instructive to contrast these predictions with the outcome of the 
direct minimization of the HRT functional, already shown in Fig. \ref{fig:scala_micro_rhocritica}. 
Such a comparison can be found in Fig. \ref{fig:micro_vs_long}, where the scaling functions obtained from the microscopic DFT 
at a few reduced temperatures $t$ in the critical region are shown to agree remarkably well with 
the predictions of the long wavelength analysis, 
provided both the scaling function for the free energy $\Psi(x)$ and 
the universal amplitude ratio $g_4^+$ are consistently
evaluated at the same reduced temperature $t$. 
\newline
The presence of strong corrections to scaling in both $\Psi(x)$ and $g_4^+$, already highlighted, induces 
strong \hbox{pre-asymptotic} effects in the critical Casimir scaling function which, 
at reduced temperatures lower than 
$10^{-3}$, is still quite far from its asymptotic limit. 
The main effect is due to the growth of the amplitude
ratio, which, as shown in Fig. \ref{fig:scalapsi_g4}, appears to reach its universal value only extremely close to 
the critical point, according to the prediction of HRT for the model of critical fluid investigated here. 
We remark that, due to the already quoted difference between the HRT estimate of the 
universal amplitude ratio $g_4^+$ and the value obtained via series expansions, the critical Casimir 
scaling function predicted by our DFT significantly differs from the one obtained in MC simulations, 
as can be seen in Fig. \ref{fig:micro_vs_long}.

\section{Conclusions and perspectives} 

We presented a novel Density Functional, based on the Weighted Density paradigm, able to 
describe classical inhomogeneous fluids in a large portion of their phase diagram, critical
point included. This is the first attempt to describe the effects of correlations induced by 
attractive interactions in confined fluids. The theory is based on the description of the
uniform system provided by the Hierarchical Reference Theory, one of the few liquid state 
approaches able to cope with long range density fluctuations. This technique, applied to
the evaluation of the effective interaction between two hard walls in a fluid, allowed 
for the investigation of the crossover between a depletion-like mechanism at high temperatures
and the critical Casimir effect emerging near the critical point. Our method 
does not rely on a long wavelength approximation and provides a complete picture of the 
solvent mediated force for any wall separation, 
displaying the presence of important non universal contributions
in the effective interaction at short distances, even in the critical region. We believe that 
this DFT will be useful in investigating other correlated systems, when density fluctuations 
are expected to play an important role. 

We showed that, at large separations, the solvent mediated force per unit surface 
between the walls decays 
exponentially on the scale of the correlation length in the whole 
portion of the phase diagram to the left of the \hbox{Fisher-Widom} line. Such a behavior 
cannot be considered as a signature of the onset of critical Casimir effect: only 
the product between the {\it amplitude} of the long range exponential tail and the
cube of the correlation length is a genuine universal quantity. 

Our microscopic approach allows for the determination of 
the universal quantities characterizing the critical Casimir effect, namely 
the scaling function $\Theta(s,y)$, both along the critical isochore ($y=0$) and in the 
\hbox{off-critical} regime. Strong corrections to scaling have been observed in the HCY fluid we 
investigated: The universal features appear to emerge only in a narrow neighborhood 
of the critical point, at least in the model we examined. 
It would be useful to compare this prediction
with numerical simulations for the HCY fluid model as well as with theoretical 
investigations of other systems, 
like the Ising model, where the correction to scaling may be weaker. 
These studies will hopefully clarify the origin of the discrepancy between the HRT
estimate of the universal amplitude ratio $g_4^+$ and the commonly accepted value. 

The approach presented in this work allows for further improvements. In our density functional,
the ideal gas term 
and the Hartree contribution to the internal energy have been treated exactly, while the 
remaining \hbox{entropy-correlation}  term has been approximated by use of a weighted density trick. 
The next step will be to treat the hard sphere term by the Fundamental Measure Theory, known
to be very accurate in dealing with excluded volume effects, limiting the weighted density   
contribution only for the residual correlation term. This adjustment is expected to 
increase the accuracy of the theory at high density, without however modifying the description of the 
universal properties of the critical Casimir effect. 

In this first application we just considered a planar geometry, whose implications for the 
phenomenon of colloidal aggregation depend upon further assumptions, 
namely the Derjaguin approximation, which however turns out to be rather inaccurate 
when the two external bodies are not very close. A natural further step will be to perform the 
functional minimization in cylindrical geometry, appropriate for dealing with two spherical 
particles thereby avoiding any \hbox{wall-to-sphere} mapping. 

Finally, a very interesting application of this formalism will be the investigation of 
\hbox{sub-critical} temperatures, where wetting phenomena are expected close to the first 
order transition boundary. The Hierarchical Reference Theory of fluids provides a consistent 
description of the full\hbox{liquid-vapor} transition line and then it appears to be the natural
starting point for the development of a microscopic theory, beyond mean field, for the 
study of phase coexistence near a wall. 



\end{document}